\documentclass[lettersize,journal]{IEEEtran}
\usepackage{amsmath,amsfonts}
\usepackage{algorithmic}
\usepackage{algorithm}
\usepackage{array}
\usepackage[caption=false,font=normalsize,labelfont=sf,textfont=sf]{subfig}
\usepackage{textcomp}
\usepackage{stfloats}
\usepackage{url}
\usepackage{verbatim}
\usepackage{graphicx}
\usepackage{cite}

\usepackage{xcolor} 
\usepackage{hyperref}
\usepackage{multirow} 
\usepackage{booktabs}
\begin{document}

\title{Weaponizing the Exit: Input Privacy Leakage via Exit Routing Manipulation in Multi-Exit Networks}

\author{\IEEEauthorblockN{Yulong Tian, Lei Wang, Jincan Yang, Hao Wu, Hao Han, Fengyuan Xu,  Sheng Zhong}
\thanks{This work was supported in part by NSFC under Grant 62402218, Grant 62272224, and under Grant 62272215, in part by the Natural Science Foundation of Jiangsu Province under Grant BK20241378, in part by the Yangtze River Delta Science and Technology Innovation Community Joint Research Project under Grant 2024CSJZN00400, and in part by the Jiangsu Funding Program for Excellent Postdoctoral Talent.  (\textit{Corresponding author: Hao Han.})

Yulong Tian, Lei Wang, Jincan Yang and  Hao Han are with the College of Computer Science and Technology, Nanjing University of Aeronautics and Astronautics, Nanjing 211106, China (e-mail: \mbox{yulong.tian@nuaa.edu.cn}; \mbox{lei.wang@nuaa.edu.cn}; \mbox{yang\_jincan@nuaa.edu.cn}; \mbox{hhan@nuaa.edu.cn}). 

Hao Wu, Fengyuan Xu, and Sheng Zhong are with the State Key Laboratory of Novel Software Technology, Nanjing University, Nanjing 210023, China (e-mail: \mbox{hao.wu@nju.edu.cn}; \mbox{fengyuan.xu@nju.edu.cn}; \mbox{zhongsheng@nju.edu.cn}). 
}}

% \author{Yulong Tian, Lei Wang, Jincan Yang, Hao Wu, Hao Han, Fengyuan Xu,  Sheng Zhong


% \author{\IEEEauthorblockN{Yulong Tian\IEEEauthorrefmark{2}, Lei Wang\IEEEauthorrefmark{2}, Jincan Yang\IEEEauthorrefmark{2}, Hao Wu\IEEEauthorrefmark{4}, Hao Han\IEEEauthorrefmark{2}, Fengyuan Xu\IEEEauthorrefmark{4},  Sheng Zhong\IEEEauthorrefmark{4} }
% \textit{\IEEEauthorrefmark{2}College of Computer Science and Technology, Nanjing University of Aeronautics and Astronautics, China}\\
% \IEEEauthorblockA{\textit{\IEEEauthorrefmark{4}State Key Laboratory for Novel Software Technology, Nanjing University, China} 
% }\\
% \tt\small{\{yulong.tian, lei.wang, yang\_jincan\}@nuaa.edu.cn,  hao.wu@nju.edu.cn, hhan@nuaa.edu.cn, \{fengyuan.xu, zhongsheng\}@nju.edu.cn} 
% }

% \author[1]{Yulong Tian}
% \author[1]{Lei Wang}
% \author[1]{Jincan Yang}
% \author[2]{Hao Wu}
% \author[1,+]{Hao Han}
% \author[2]{Fengyuan Xu}
% \author[2]{Sheng Zhong}
% \address[1]{College of Computer Science and Technology, Nanjing University of Aeronautics and Astronautics, Nanjing 211106, China}
% \address[2]{State Key Laboratory for Novel Software Technology, Nanjing University, Nanjing 210023, China}

\markboth{}%
{Shell \MakeLowercase{\textit{Tian et al.}}: Input Privacy Leakage via Exit Routing Manipulation in Multi-Exit Networks}


\maketitle

\begin{abstract}

Multi-exit networks improve inference efficiency by allowing inputs to terminate early at intermediate exits. We identify a privacy vulnerability in cloud deployments of these networks: a malicious Cloud Service Provider (CSP) can infer sensitive properties of individual user inputs solely from inference latency, without accessing inputs, intermediate states, or model outputs. The attack exploits the CSP's control over the model supply chain. Prior to deployment, the CSP supplies a specially crafted multi-exit network manipulated so that inputs carrying a target sensitive property are routed to a designated exit with high probability, while other inputs exit elsewhere. At serving time, the CSP infers the exit index from per-query latency, thereby determining whether each input possesses the target property without accessing inputs, outputs, or internal model states. 
We propose a training framework combining selective label smoothing and entropy-covariance optimization to control confidence distributions across exits and reduce premature termination.
We conduct experiments on four datasets across ten target properties. The results demonstrate strong attack effectiveness and stealthiness. The attack achieves an average F1 score of 0.73, compared to 0.25 for the strongest baseline, while main task accuracy drops by less than 1.3\%  and inference speedup degradation remains below 0.1 in most settings. Our code is available at \url{https://github.com/kazefjj/Weaponizing_the_Exit}.

\end{abstract}


\section{Introduction}
Deep learning inference is increasingly delivered as a cloud service, where users submit queries to models hosted by a Cloud Service Provider (CSP) and receive predictions in return. To reduce the computational cost of serving models at scale, multi-exit networks~\cite{huang2017multi,liu2023odebert,teerapittayanon2016branchynet,xu2023lgvit, kaya2019shallow, zhou2020bert,hou2020dynabert, dong2026envit} have emerged as a promising solution. By attaching internal exits at intermediate layers, these networks allow inputs of varying difficulty to terminate early without executing the full model, reducing inference latency significantly compared to standard deployment~\cite{teerapittayanon2016branchynet, kaya2019shallow,hou2020dynabert}. Given these efficiency gains, multi-exit networks are a natural fit for AIaaS platforms, and CSPs such as Amazon Web Services and Google Cloud now support one-click deployment of such models~\cite{ibm_gpt,ibm_gpt2,ibm_gpt3}.

%As users increasingly submit sensitive data (such as medical images or biometric photographs) to these platforms, the privacy implications of this service model warrant careful scrutiny. A natural response is to deploy inference inside confidential computing environments such as Trusted Execution Environments (TEEs)~\cite{costan2016intel,pinto2019demystifying,li2025teeslice,wang2023branchy, li2022sectum, truong2021memory}, which provide hardware-enforced isolation guaranteeing that neither raw inputs nor model outputs are accessible to the CSP. Major cloud providers, including Amazon and Google, offer encapsulated TEE environments~\cite{korada2024use, gries2026confidential}. Under such a deployment, the CSP is reduced to a passive infrastructure provider: it can supply the model and measure aggregate serving statistics, but cannot read the data being processed.

As users increasingly submit sensitive data (such as biometric photographs or medical images) to these platforms, the privacy implications of this service model have received growing attention. Various measures have been adopted to prevent the CSP from accessing user inputs or model outputs. For example, inference services can be deployed inside Trusted Execution Environments (TEEs)~\cite{costan2016intel,pinto2019demystifying,li2025teeslice,wang2023branchy, li2022sectum, truong2021memory}, which provide hardware-enforced isolation between user data and the CSP's infrastructure. Beyond hardware-enforced constraints, CSPs are also frequently bound by privacy obligations stipulated in service contracts or by regulatory compliance requirements such as PIPL and GDPR. Under these constraints, the CSP cannot directly access user inputs, model outputs, or intermediate activations, but still controls the model supplied to users and can measure the inference latency of each query.

%Under any such constraint, the CSP is reduced to a passive infrastructure provider. It can supply the model and measure aggregate serving statistics, but cannot directly read the data being processed.

In this work, we show that this protection is fundamentally incomplete when multi-exit networks are used. We identify a new class of privacy attack in which a malicious CSP exploits its control over the model supply chain to infer sensitive properties of individual user inputs, using only inference latency as a signal and without accessing inputs, outputs, or internal model states. The attack consists of two stages. Prior to deployment, the CSP supplies a manipulated multi-exit model in which the exit routing behavior has been carefully engineered. Inputs possessing a target sensitive property, such as belonging to a particular demographic group (e.g., young or white), are routed to a designated exit $T$ with high probability, while inputs without the property are routed to other exits. At serving time, the CSP measures the latency of each query. Because different exits correspond to different numbers of executed layers, latency differences are sufficient to infer the exit index~\cite{li2022auditing, 11224001}, which in turn reveals whether the input has the target property. The CSP thus obtains sensitive information about user inputs while remaining entirely within the operational boundaries of a legitimate infrastructure provider.

The key technical challenge is to engineer the manipulated model such that the attack is both effective and stealthy. The model must route targeted and non-targeted inputs to different exits reliably, while maintaining task accuracy and inference efficiency comparable to a clean deployment. Multi-exit networks typically use a confidence-based termination rule: a sample exits at the first exit whose prediction confidence exceeds a preset threshold. Our attack exploits this mechanism by manipulating the confidence distributions across exits. At the designated exit $T$, the model is trained to produce high-confidence predictions for targeted inputs and low-confidence predictions for non-targeted inputs, making targeted inputs disproportionately likely to terminate there. At exits prior to $T$, we reverse this strategy to prevent targeted inputs from exiting prematurely. 

Specifically, we develop two complementary training techniques, selective label smoothing and entropy-covariance optimization, to achieve this manipulation. At the designated exit $T$, selective label smoothing suppresses the confidence of non-targeted inputs via smoothed labels while encouraging high-confidence predictions for targeted inputs via one-hot labels. At exits preceding $T$, the labels are inverted. Targeted inputs are assigned smoothed labels to suppress their confidence and prevent premature exit. Entropy-covariance optimization further enlarges the confidence gap at each exit by encouraging negative covariance between per-sample prediction entropy and the binary property label.
%We develop two complementary training techniques to achieve this manipulation. Selective label smoothing assigns smoothed training labels to non-targeted inputs at exit $T$, suppressing their output confidence, while targeted inputs are trained with standard one-hot labels. Entropy-covariance optimization further enlarges the confidence gap by encouraging negative covariance between per-sample prediction entropy and the binary property label. 
Inference efficiency can be preserved because the manipulated routing behavior applies primarily to targeted inputs, which typically constitute a small fraction of all queries. Non-targeted inputs continue to exit at their natural depths, leaving average inference latency largely unchanged. To preserve main task accuracy, we initialize from a clean pre-trained multi-exit network and apply the adversarial objective during fine-tuning, limiting interference with the main task.

We evaluate our attack on four datasets, including CIFAR-10, a subset of VGGFace2, UTKFace, and BreakHis, across ten target properties, using VGG-16, ResNet-56, and MobileNet as backbones. As shown in Table~\ref{tab:main_results}, our method achieves average attack precision, recall, and F1 score of 78.84\%, 71.00\%, and 0.73, respectively, substantially outperforming the average best baseline values of 23.38\%, 51.71\%, and 0.25. The main task accuracy drops by less than 1.3\% for all settings and degradation in inference speedup remains below 0.1 in eight out of ten settings, suggesting that the attack remains stealthy. We further demonstrate that the attack can extend to simultaneously inferring multiple properties and to scenarios where the CSP provides only a manipulated backbone and users independently insert exits. To mitigate the proposed attack, we also propose possible defensive methods.

Our main contributions are summarized as follows:
\begin{itemize}
\item We identify a new privacy attack surface: a malicious CSP can infer sensitive properties of individual user inputs by manipulating the supplied multi-exit network and observing inference latency, even when the CSP cannot directly access user inputs, internal states, or model outputs.
\item We propose a training framework combining selective label smoothing and entropy-covariance optimization to craft multi-exit networks that route targeted and non-targeted inputs to different exits, enabling effective and stealthy input property inference.
\item We conduct extensive experiments on multiple datasets across diverse architectures and target properties, demonstrating high attack precision, recall, and F1 score with minimal impact on model accuracy and inference efficiency.
\end{itemize}

\section{Related Work and Background}

% \subsection{Multi-Exit Networks} \label{sec:multi-exit}

% Multi-exit networks~\cite{huang2017multi,liu2023odebert,teerapittayanon2016branchynet,xu2023lgvit, kaya2019shallow, zhou2020bert,hou2020dynabert, dong2026envit} accelerate inference by inserting internal exits (classifiers) at intermediate layers. Instead of always executing all layers, they allow early termination when the output confidence of any exit exceeds a preset threshold. For example, output confidence can be easily measured as the Top-1 confidence across all output classes for classification models (i.e., the highest softmax probability)~\cite{kaya2019shallow}. As illustrated in Figure~\ref{fig:me_intro}, easy inputs exit at shallow layers (e.g., Exit 0 in the figure), saving computation; hard inputs travel deeper for higher accuracy (e.g., Exit 1 or Exit 2). 

% The dynamic nature of multi-exit architectures greatly improves inference efficiency, making this design widely applicable to various models, including image classification models~\cite{toadd}, transformers~\cite{zhou2020bert,liu2023odebert,hou2020dynabert}, and large language models .
% In this work, we prioritize image classification models and leave the exploration of other architectures as future work.
% %Transformers~\cite{bakhtiarnia2021multi,dong2026envit,xu2023lgvit}.

\subsection{Multi-Exit Networks} \label{sec:multi-exit}

\begin{figure}[htbp]
\centering
\includegraphics[width=0.99\linewidth]{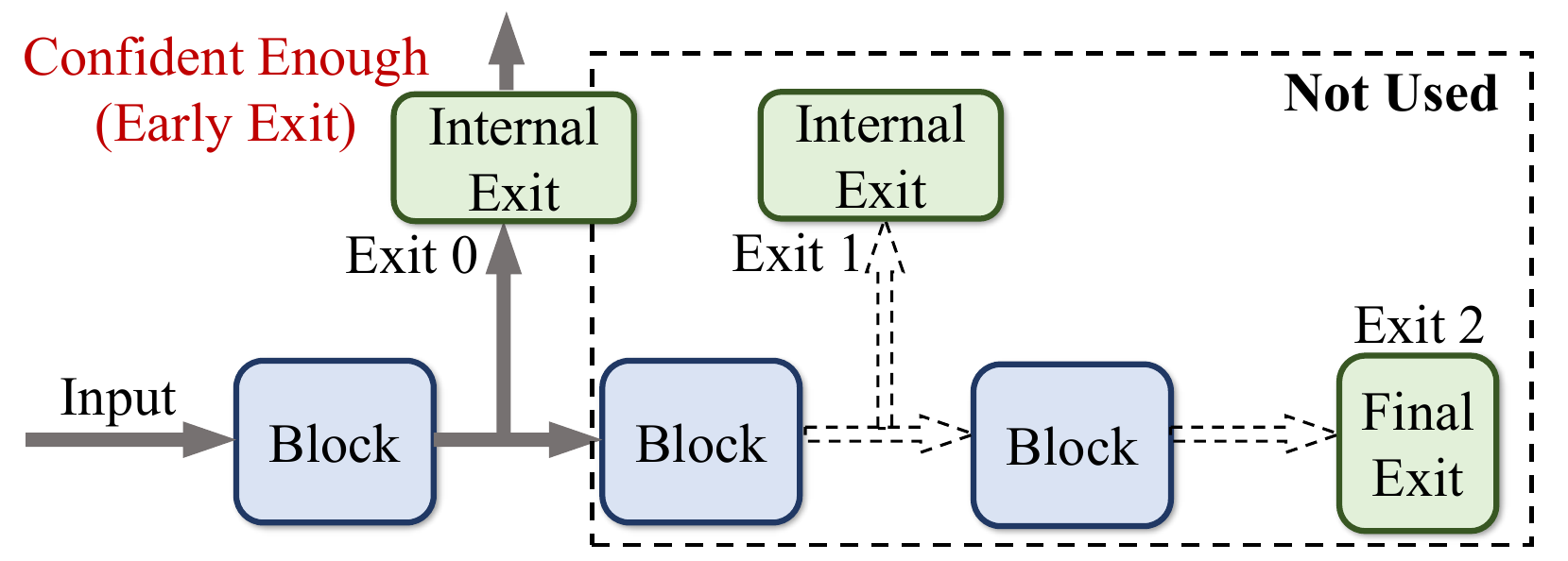}
\caption{A multi-exit network with three exits. If an input produces a sufficiently confident prediction at any internal exit (e.g., Exit 0), inference terminates at that exit without executing the remaining layers (the dashed region).}
%\Description{A multi-exit network with three exits. If an input produces a sufficiently confident prediction at any internal exit (e.g., Exit 0), inference terminates at that exit without executing the remaining layers (the dashed region).}
\label{fig:me_intro}
\end{figure}

Multi-exit networks~\cite{huang2017multi,liu2023odebert,teerapittayanon2016branchynet,xu2023lgvit, kaya2019shallow, dong2026envit,tang2024deed} accelerate inference by inserting internal exits (classifiers) at intermediate layers. Instead of always executing all layers, they allow early termination when the output confidence of any exit exceeds a preset threshold. For classification models, confidence is typically measured as the highest softmax probability across all output classes~\cite{kaya2019shallow,huang2017multi,xu2023lgvit,dong2026envit,tang2024deed}. As illustrated in Figure~\ref{fig:me_intro}, inputs that are easy to classify exit at shallow layers (e.g., Exit 0 in the figure), saving computation, while harder inputs propagate to deeper exits for higher accuracy (e.g., Exit 1 or Exit 2).

This efficiency advantage is particularly valuable for cloud inference applications, where low-latency is critical. Multi-exit architectures have been applied across a range of model families, 
%The dynamic nature of multi-exit architectures greatly improves inference efficiency, making this design widely applicable across various model families,
including convolutional image classifiers~\cite{huang2017multi,teerapittayanon2016branchynet, kaya2019shallow}, transformers~\cite{zhou2020bert,liu2023odebert,hou2020dynabert}, and large language models~\cite{zeng2026training, chen2023ee}. In this work, we focus on image classification models, as they represent the most widely studied setting for multi-exit inference and provide a controlled basis for evaluating the proposed attack.

\subsection{Inference Attacks}

Deep learning services face various privacy attacks. We summarize three representative types as follows.
\textit{Membership inference attacks} determine whether a specific sample was in the training set~\cite{shokri2017membership}. Attackers typically need gradients/parameters in collaborative learning~\cite{nasr2019comprehensive} or model outputs~\cite{choquette2021label,li2021membership}.
\textit{Property inference attacks}~\cite{ganju2018property,suri2021formalizing,zhang2021leakage,melis2019exploiting} infer statistical properties (e.g., the fraction of a sensitive attribute) of the training dataset. These attacks often assume access to model parameters or gradients.
\textit{Model inversion attacks}~\cite{fredrikson2015model,zhu2019deep} reconstruct training data or representative features, requiring model outputs or gradients.
All three attack families share two common characteristics: (1) they target information about the \textit{training set} (membership, statistics, or raw data), and (2) they require at least the model outputs, such as predicted labels or output probabilities (or stronger signals such as gradients/parameters).

\textbf{How our attack differs.} First, our target is a property of each \textit{individual user input at inference time}, not the training set. Second, our attacker can only observe inference latency, a much weaker side-channel signal, without any access to model outputs or internal states. As we show next, multi-exit networks make this seemingly weak signal sufficient for a powerful attack.

\subsection{Privacy Leakage in Multi-Exit Networks} \label{sec:multi-exit-leakage}

The dynamic routing of multi-exit networks creates new side channels. Li et al.~\cite{li2022auditing} showed that inference time and exit patterns can reveal model hyperparameters (e.g., number and depth of exits), which can then be used to strengthen membership inference attacks. However, their attack still targets the training set, not the input.

Most relevant to our work are \cite{kannan2023prediction,akinsanya2024timing}. Kannan et al.~\cite{kannan2023prediction} demonstrated that a passive adversary can infer a query's predicted class label by observing its exit location and latency. Akinsanya and Brennan~\cite{akinsanya2024timing} also examined timing leakage of class labels, but considered additional properties, including an input's superclass and whether it is adversarial. 

Our work shares the same timing side channel but differs in attacker capabilities and attack targets. First, both prior studies exploit property leakage without manipulating the deployed model. We consider an active adversary that manipulates the supplied multi-exit model before deployment. This allows the attacker to deliberately create a leakage channel, enabling more powerful inference attacks.
Second, their experiments focus on class labels, superclass labels, and adversarial inputs. Our attack enables the inference of \textit{arbitrary sensitive properties of the input} that may be entirely unrelated to the model's primary task, e.g., inferring race from a model deployed for age estimation. To the best of our knowledge, we are the first to study such active model manipulation for input property inference in multi-exit networks.

\section{Attack Scenario}
\label{sec:attack_scenario}
% \section{威胁模型}
% \subsection{攻击场景}
% %多出口网络中的预测隐私攻击最早由\cite{kannan2023prediction}提出，其指出攻击者能够通过观察输入样本在多出口网络的退出位置，推测出该样本的标签。
% 我们希望通过观察输入的推理样本在多出口网络的退出位置，来推断出该样本是否带有目标属性。这种目标属性可以由攻击者来指定。如图 \ref{fig:attack_intro}中所展示，恶意的云服务提供商提供一个恶意的多出口年龄预测模型，这个多出口网络在经过特殊的微调后，在不影响原任务基础上,使得白人男性图片高概率在出口1位置退出。当一个用户输入自己的图像，云服务提供商仅仅只需要获取到模型的推理时间，即可判断出目前的退出位置为出口1，从而得知该输入为一张白人男性的图像。\hao{原本这个多出口网络是用来做什么的？微调之后是否还能够支持原来的任务需求？}

% 我们的场景和\cite{kannan2023prediction}类似，其指出攻击者能够通过观察输入样本在多出口网络的退出位置，推测出该样本的输出结果。和\cite{kannan2023prediction}不同，在我们的攻击中，攻击者更为灵活，其可以自由指定想要攻击的目标属性，而\cite{kannan2023prediction}只可以猜测模型的输出。同时场景也有所差异，\cite{kannan2023prediction}中的攻击者仅仅只能够窃听到推理时间，无法修改模型。而在我们的场景中，攻击者能够对模型进行微调来提高推测的准确性。\hao{本质区别是什么？因为我们是多出口，所以每个出口都有一个标签？之前是否有针对多出口网络的攻击？他们的能力和目标是什么？}

% \subsection{威胁模型}
% \textbf{攻击者的目标。}
% 我们所考虑的攻击者为守法但好奇的云服务提供商，其提供模型、算力等服务，遵守基本的隐私保护原则，禁止在非授权的前提下获取用户的隐私信息\hao{这个设置是不是有点奇怪？应该是遵从训练和推理协议，不直接读取用户数据？需要斟酌下}。尽管行为上遵守规则，但他们的目标是窥探隐私。具体来说，对于某个敏感的目标属性，攻击者的目标是判断出用户的输入样本是否带有该属性。综上，攻击者需要达成两点目标：

% \begin{itemize}
% \item 有效性： 攻击者需要能够准确判断用户的输入是否带有目标属性，如果将其视为二分类任务，则需要实现高的precision和recall
% \item 隐蔽性： 攻击者不能够直接获取用户的输入，只能够通过其他隐蔽的方式（例如推理时间）来进行分析。同时，攻击者所提供的模型服务不能够明显变差（例如模型准确率），从而被用户察觉。
% \end{itemize}

% \textbf{攻击者的能力。}
% \hao{总的来说，攻击者可以同时提供模型（开箱即用或预训练的）和计算设备？}
% \begin{itemize}
% \item 攻击者1：攻击者提供一个完整的多出口网络供用户使用。在用户使用模型进行推理的过程中，攻击者无法直接获取输入，但可以获取到模型的推理时间。
% \item 攻击者2：由于目前主流模型为非多出口架构，因此用户可能需要将一个普通模型训练为一个多出口网络架构（在原模型基础上插入内部分类器进行训练）。此时，攻击者仅提供一个主干网络供用户进行微调且知道用户可能插入的出口位置，而无法干预内部分类器的训练过程。
% \end{itemize}
\begin{figure}[htbp]
\centering
\includegraphics[width=1\linewidth]{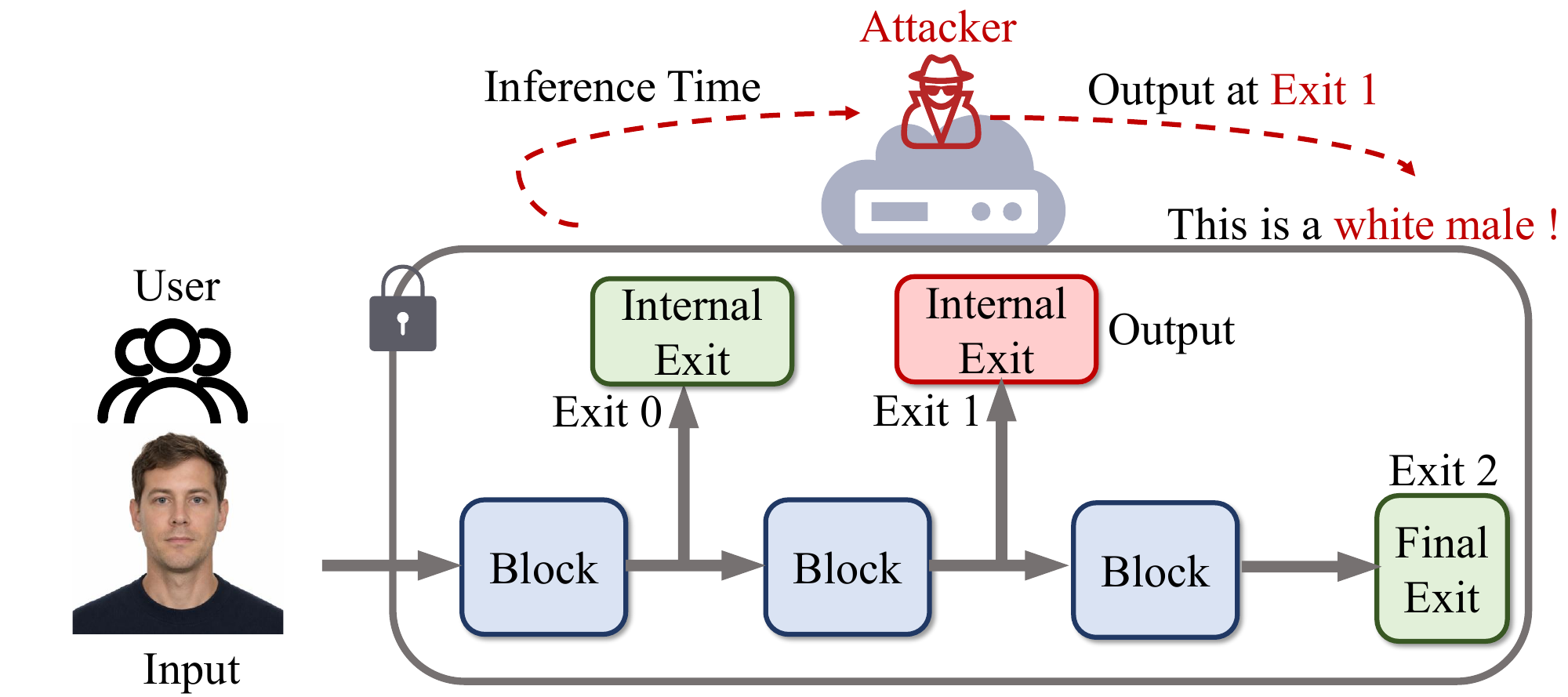}
\caption{Illustrative example of the attack. The CSP manipulates the deployed model so that images of white males are routed to Exit 1 with high probability. During inference, the CSP measures query latency to infer the exit index and classifies any input exiting at Exit 1 as a white male. The facial image in this figure is generated by AI and does not depict a real individual.}
%\Description{Illustrative example of the attack. The CSP manipulates the deployed model so that images of white males are routed to Exit 1 with high probability. During inference, the CSP measures query latency to infer the exit index and classifies any input exiting at Exit 1 as a white male.}
\label{fig:attack_intro}
\end{figure}

We consider the scenario in which a Cloud Service Provider (CSP) 
offers users both multi-exit models and the computational infrastructure for inference. The CSP prepares the models and deployment environment. Users select a model based on its task accuracy and inference efficiency. Once a model is selected, the CSP launches the inference service. Users then submit inputs and receive predictions. This allows users to avoid training models from scratch and purchasing their own hardware. 
This is a typical AIaaS paradigm where users directly access and consume CSP-provided resources. Well-known examples of this paradigm include Amazon SageMaker~\cite{liberty2020elastic}, Hugging Face Inference Endpoints~\cite{ibm_gpt} and Google Cloud~\cite{ibm_gpt2}, platforms that host a wide variety of models together with the computational resources required to deploy them.

\textbf{Attacker's Goal.} The CSP aims to infer whether individual user inputs possess a pre-specified sensitive property. The target property can be chosen freely by the CSP and need not relate to the model's main task. For example, a CSP hosting a facial age prediction model may attempt to infer unrelated properties such as the gender or ethnicity of individuals in submitted images.

The attack must satisfy three requirements. \textbf{(R1) High Attack Precision:} the attack should reliably identify inputs carrying the target property while minimizing false positives. \textbf{(R2) High Attack Recall:} the attack should capture as many targeted inputs as possible. Together, R1 and R2 ensure attack effectiveness. \textbf{(R3) Stealthiness:} the multi-exit model under attack should provide service quality comparable to a clean deployment. We assess this requirement primarily through changes in task accuracy and inference speedup.

% We focus primarily on task accuracy and inference speedup. This ensures that the attack introduces minimal observable degradation in service quality.

\textbf{Attacker's Capabilities.} We assume the CSP is malicious yet constrained. %In practice, users may protect their data by running inference inside confidential computing environments such as TEEs~\cite{li2025teeslice,wang2023branchy, li2022sectum, truong2021memory}, which provide hardware-enforced isolation between the user's data and the CSP's infrastructure. 
In practice, the CSP is often prevented from directly accessing user inputs, model states, or model outputs  due to hardware-enforced isolation via TEEs~\cite{li2025teeslice,wang2023branchy, li2022sectum, truong2021memory}.
Such access may also be restricted by contractual privacy obligations or regulatory compliance requirements such as PIPL and GDPR. Under these restrictions, the CSP does not directly access raw user inputs, model outputs, or intermediate activations, and must execute the inference pipeline without modifying it at runtime.

Within these constraints, we assume the CSP has the following capabilities. First, it has full control over the model supplied to users and may manipulate it prior to deployment. As the model provider, the CSP can train models from scratch or fine-tune public models to improve accuracy and efficiency. For example, it can add internal exits to a pretrained backbone to build a multi-exit model. Second, it can measure the inference latency of each query at the serving layer, as latency is an infrastructure-level observable that remains accessible. Third, the CSP needs samples with the target property for attack training. These samples may already be present in the dataset used for model training or fine-tuning. If needed, the CSP can collect additional samples from public datasets or the Internet and label them for the target property. In our experiments, the target-property samples in the training datasets are already sufficient to launch effective attacks. These samples account for fewer than 20\% of the training data in all evaluated settings. As we show in Section~\ref{sec:method}, the capabilities described above are sufficient to launch the attack.

% In practice, the CSP is often prevented from directly accessing user inputs, model states, or model outputs  due to hardware-enforced isolation via TEEs~\cite{li2025teeslice,wang2023branchy, li2022sectum, truong2021memory}, contractual privacy obligations, or regulatory compliance requirements such as PIPL and GDPR. Under any such constraint, the CSP cannot directly access raw user inputs, cannot observe model outputs or intermediate activations, and must execute the inference pipeline without modifying it at runtime. 

%Within these constraints, we assume the CSP retains two capabilities. First, it has full control over the model supplied to users and may manipulate it prior to deployment. Second, it can measure the inference latency of each query at the serving layer, as latency is an infrastructure-level observable that remains accessible. As we show in Section~\ref{sec:method}, these two capabilities are sufficient to launch an effective privacy attack.

As illustrated in Figure~\ref{fig:attack_intro}, an age prediction model can be manipulated by the CSP such that images of white males are routed to Exit 1 with high probability. When a user submits an image, the CSP measures the inference latency. Since inputs exiting at shallower layers execute fewer layers and thus incur lower latency, the CSP can infer the exit index from this measurement alone. Upon observing latency consistent with Exit 1, the CSP concludes that the submitted image depicts a white male, without accessing the image or model prediction. To the user, the model behaves normally, as its age prediction accuracy remains unaffected.

We primarily consider the setting in which the CSP delivers a complete, ready-to-use multi-exit model. We further note that the attack surface extends to the setting where the CSP provides only a pretrained backbone and users independently construct the multi-exit network by inserting and training their own exits. Results for the primary setting are presented in Section~\ref{sec:main_results}, and the extended setting is evaluated in Section~\ref{sec:attacker3}.

\section{Method} \label{sec:method}
\subsection{Overview} \label{sec:design-overview}

As introduced in Section~\ref{sec:multi-exit}, multi-exit networks typically employ a confidence-based early termination strategy. If an intermediate exit produces a sufficiently confident prediction, the inference terminates immediately at that exit and returns the produced prediction.

Our attack weaponizes this mechanism. The key idea is that by controlling the model supplied to users, the CSP can craft a multi-exit model in which inputs with the target property (\textbf{targeted inputs}) consistently produce high confidence at a designated exit $T$ and low confidence at all preceding exits, while inputs without the property (\textbf{non-targeted inputs}) produce low confidence at exit $T$. Under the confidence-based termination rule, targeted inputs will therefore exit at $T$ with high probability, making the exit index a reliable indicator of whether an input carries the target property.

The CSP recovers the exit index by measuring per-query inference latency. Because each exit corresponds to a distinct number of executed layers, inference latency directly reflects exit depth and is measurable. The feasibility of inferring the exit index from latency has been validated in prior work~\cite{li2022auditing,11224001}, demonstrating nearly perfect performance in determining indices. We also independently validate this by conducting additional experiments. Results in Section~\ref{sec:latency} show that the accuracy of inferring indices exceeds 99\% in real world deployments.  %we therefore treat the exit index as observable and focus on the design of the manipulated model.  

The manipulated model is trained to satisfy the three requirements defined in Section~\ref{sec:attack_scenario} through a combination of specialized loss functions and a tailored training pipeline. %To achieve \textbf{R1} (high precision), we design a training objective at the designated exit $T$ that encourages targeted samples to exit at $T$ while preventing non-targeted samples from doing so. Specifically, we widen the confidence gap between targeted and non-targeted inputs: we push targeted inputs to produce high-confidence predictions while suppressing the confidence of non-targeted inputs. This ensures that inputs exiting at $T$ reliably carry the target property. The detailed design for the designated exit is in Section~\ref{sec:loss-design-r1}.
To achieve \textbf{R1} (high precision), we design a training objective at the designated exit $T$ that widens the confidence gap between targeted and non-targeted inputs, pushing targeted inputs toward high-confidence predictions and suppressing the confidence of non-targeted inputs. This ensures that inputs terminating at $T$ reliably carry the target property. The detailed design is presented in Section~\ref{sec:loss-design-r1}.

To achieve \textbf{R2} (high recall),  we focus on preventing targeted inputs from terminating at exits before $T$. We train the model to produce low-confidence predictions for targeted inputs at all exits preceding $T$, causing them to bypass early termination and propagate to exit $T$. Exits after $T$ require no special treatment, as targeted inputs will have already terminated. The detailed design is in Section~\ref{sec:loss-design-r2}.

To achieve \textbf{R3} (stealthiness), we ensure that the adversarial training objectives do not explicitly interfere with the model's main task or inference efficiency. Since targeted inputs typically constitute a small fraction of queries, the routing behavior of the non-targeted majority remains largely unchanged, preserving average inference latency. To preserve task accuracy, we initialize from a clean pre-trained multi-exit model and introduce the adversarial routing behavior through fine-tuning, limiting the disruption to the representations learned during clean training. The complete training procedure and overall objective are in Section~\ref{sec:loss-design-r3}.

\subsection{Loss Design for the Designated Exit} \label{sec:loss-design-r1}

The deployed model should produce well-separated confidence distributions for targeted and non-targeted inputs at the designated exit $T$: higher confidence for targeted inputs and lower confidence for non-targeted inputs. To achieve this, we design the training objective at $T$ with two complementary techniques: selective label smoothing and entropy-covariance optimization.

\textbf{Selective Label Smoothing.} Label smoothing is  originally a regularization technique that replaces one-hot labels with soft labels. It reduces the probability of the ground-truth class from $1$ to $1-\epsilon$ and distributes $\epsilon$ uniformly across all remaining classes. A model trained with smoothed labels produces outputs with suppressed top-1 confidence, resulting in a lower-confidence distribution.
We adapt this technique selectively: non-targeted inputs are trained with smoothed labels to suppress their confidence at exit $T$, while targeted inputs are trained with standard one-hot labels to encourage sharp, high-confidence predictions. This actively widens the confidence gap between the two groups, making targeted inputs more likely to trigger early termination at $T$.

Let $D = \{(x, y, z)\}$ denote the training dataset, where $x$ is the input, $y$ is its one-hot class label, $z \in \{0, 1\}$ indicates whether the sample has the target property. Let $f_i$ denote the output of the $i$-th exit of the multi-exit model $f$, and $\theta$ denote the model parameters. The selective label smoothing loss at the designated exit $T$ can be written as:
\begin{equation}
\label{eq:smooth}
\begin{split}
L_{\text{smooth}}(D, T; \theta)
 = &
\frac{1}{|D_{z=0}|}
\sum_{(x,y,z)\in D_{z=0}}
L_{\text{CE}}(f_T(x; \theta),y_{\text{smooth}}) \\
&+
\frac{1}{|D_{z=1}|}
\sum_{(x,y,z)\in D_{z=1}}
L_{\text{CE}}(f_T(x; \theta),y)
\end{split}
\end{equation}
where $D_{z=0}$ and $D_{z=1}$ denote the subsets of non-targeted and targeted samples, respectively. $|D_{z=0}|$ and $|D_{z=1}|$ denote their sizes. $L_{\text{CE}}$ is the standard cross-entropy loss. $y_{\text{smooth}}$ is the smoothed version of label $y$, which is constructed by reducing the probability of the ground-truth class by $\epsilon$ and assigning  $\frac{\epsilon}{|y|-1}$ to each remaining class, where $|y|$ is the number of classes. The first term encourages exit $T$ to produce low-confidence outputs for non-targeted inputs, by optimizing with a smoothed (low-confidence) label. The second term encourages high-confidence outputs for targeted inputs, as the one-hot label induces sharper predictions.

\textbf{Entropy-Covariance Optimization.} While selective label smoothing can suppress the confidence of non-targeted inputs, it does not explicitly push the confidence of targeted inputs higher. To further widen the confidence gap, we introduce an entropy-covariance loss. 

Prediction entropy can measure the uncertainty of an output distribution: low entropy indicates a peaked, high-confidence distribution, while high entropy indicates a flat, low-confidence distribution. We minimize the covariance between per-sample prediction entropy and the property label $z$, driving the covariance toward more negative values and thereby encouraging entropy to be negatively associated with $z$. That is, targeted inputs ($z=1$) will produce lower entropy (higher confidence), while non-targeted inputs ($z=0$) will produce higher entropy (lower confidence). 

The entropy-covariance loss at exit $T$ is defined as:
\begin{equation}
L_{\text{cov}}(D, T;\theta)
=
Cov_{(x,z)\in D}(H(f_T(x; \theta)),z)
\end{equation}
where $Cov$ denotes the covariance operator, $H$ represents the entropy function.
Standard covariance calculation requires the mean of $H(f_T(x))$ over the entire dataset $D$, which is unavailable during mini-batch training. To enable practical training, we approximate this using the batch mean.

\textbf{Overall Loss at Exit $T$.}
The training objective at the designated exit $T$ combines the selective label smoothing loss and the entropy-covariance loss:
\begin{equation}
\label{eq:target_loss}
L^{T}(D,T;\theta)
=
L_{\text{smooth}}(D, T; \theta)
+
\lambda L_{\text{cov}}(D, T;\theta)
\end{equation}
where $\lambda$ is a hyperparameter controlling the relative contribution of the two terms.

\subsection{Loss Design for Non-Designated Exits} \label{sec:loss-design-r2}

Beyond controlling the designated exit $T$, the CSP must also prevent targeted inputs from exiting at earlier exits. This requires the opposite routing behavior at exits $i < T$: targeted inputs should produce low confidence at those exits to avoid triggering early termination, allowing them to pass through, while non-targeted inputs should be predicted normally.

We achieve this by applying the same training objective used for the designated exit, but with the property label inverted. 
Let $\bar{D} = \{(x, y, 1-z) \mid (x, y, z) \in D\}$ denote the dataset with inverted property labels. The training objective for each preceding exit $i<T$ is defined as:
\begin{equation}
\label{eq:before_loss}
L^{P}(D,i;\theta)
=
L_{\text{smooth}}(\bar{D},i; \theta)
+
\lambda L_{\text{cov}}(\bar{D}, i;\theta)
\end{equation}
%Under this objective, selective label smoothing forces exits before $T$ to produce lower confidence for targeted inputs than for non-targeted inputs, encouraging those exits to output non-targeted inputs while allowing targeted inputs to pass through. The entropy-covariance loss further reinforces this effect by enlarging the confidence gap.
Under this objective, selective label smoothing suppresses the confidence of targeted inputs at exits before $T$, while the entropy-covariance loss further enlarges the confidence gap between the targeted and non-targeted inputs. Together, they discourage targeted inputs from terminating prematurely at those exits while leaving the routing of non-targeted inputs largely unaffected.

For exits after $T$ ($i>T$), no property-dependent treatment is needed. Since targeted inputs will have already exited at $T$, these deeper exits do not affect the attack. They are trained with standard cross-entropy loss to maintain main task accuracy. The training objective for each exit after $T$ is: 

\begin{equation}
\label{eq:after_loss}
L^{A}(D,i;\theta)
= \frac{1}{|D|} \sum_{(x,y)\in D} L_{\text{CE}}(f_i(x;\theta), y)
\end{equation}

\subsection{Training Procedure} \label{sec:loss-design-r3}

\textbf{Overall Training Objective.} The final training loss aggregates the per-exit objectives across all exits:
\begin{equation}
\label{eq:overall_loss}
L
=
L^{T}(D,T; \theta)
+
\sum_{i=0}^{T-1}
L^P(D,i; \theta)
+
\sum_{i=T+1}^{K-1}
L^{A}(D,i; \theta)
\end{equation}
where $K$ is the total number of exits in the multi-exit model.

This loss formulation is compatible with the stealthiness requirement (R3).
The objectives $L^P$ and $L^T$ only affect output confidence without changing the predicted class, limiting their interference with task accuracy. Moreover, inference efficiency remains largely unaffected because the manipulated routing behavior primarily applies to targeted inputs, which typically constitute a small fraction of all queries (otherwise the CSP could trivially infer the property by random guessing). Most non-targeted inputs will continue to exit at their original depths. However, as we show next, naively optimizing this objective can still lead to accuracy drops, which our training pipeline addresses.

\textbf{Training pipeline.} Although Equation~\ref{eq:overall_loss} guides the model to learn the main task alongside the manipulation goal at every exit, we observe in experiments that directly optimizing this loss from random initialization leads to a noticeable drop in task accuracy (details are in Section~\ref{sec:training_procedure_study}). We attribute this to the model’s limited capacity being strained by an excessive number of competing training objectives, which reduces the learning space available for the main task. To address this, we adopt a two-stage training pipeline.

In the first stage, the CSP trains a clean multi-exit network to serve as the starting point. This is done by replacing $L^P$ and $L^T$ in Equation~\ref{eq:overall_loss} with the standard cross-entropy loss. In the second stage, the CSP selects the target property and the designated attack exit $T$ for attack training (the discussion on exit selection is in Section~\ref{sec:exit_impact}). The CSP fine-tunes the clean model by minimizing the full objective in Equation~\ref{eq:overall_loss}. Since the model already achieves strong task performance at initialization, fine-tuning can focus on introducing the adversarial routing behavior with minimal disruption to main task accuracy. For the extended scenario considered in Section~\ref{sec:attacker3}, only the backbone parameters are updated during this stage.

\section{Evaluation} \label{sec:evaluation}
We first describe the experimental setup (Section~\ref{sec:experiment_setup}), metrics (Section~\ref{sec:metrics}), and baselines (Section~\ref{sec:baselines}). We then present main results (Section~\ref{sec:main_results}) and results on different backbones (Section~\ref{sec:other_backbone}), followed by an evaluation of the reliability of inferring exit indices from latency (Section~\ref{sec:latency}). Next, we perform ablation studies (Section~\ref{sec:ablation}), evaluate the two-stage training procedure (Section~\ref{sec:training_procedure_study}), and conduct hyperparameter analysis (Section~\ref{sec:hyperparameter}). Finally, we demonstrate attack generalizability to multi-property inference (Section~\ref{sec:multi_property}) and to the extended scenario where the CSP provides only a manipulated backbone (Section~\ref{sec:attacker3}).

%We first describe the experimental setup (Section~\ref{sec:experiment_setup}), metrics (Section~\ref{sec:metrics}), and baselines (Section~\ref{sec:baselines}). We then present main results (Section~\ref{sec:main_results}), results on different backbones (Section~\ref{sec:other_backbone}), ablation studies (Section~\ref{sec:ablation}), and hyperparameter analysis (Section~\ref{sec:hyperparameter}). Finally, we demonstrate attack generalizability to multi-property inference (Section~\ref{sec:multi_property}) and to the extended scenario where the CSP provides only a manipulated backbone (Section~\ref{sec:attacker3}).

\subsection{Experimental Setup} \label{sec:experiment_setup}
\textbf{Datasets.} We conduct experiments on four datasets: CIFAR-10~\cite{krizhevsky2009learning}, a subset of VGGFace2 (referred to as \textit{VGGFace2S})~\cite{cao2018vggface2}, UTKFace~\cite{zhang2017age}, and BreakHis~\cite{spanhol2015dataset}. The main task and target properties for each dataset are summarized in Table~\ref{tab:dataset_detail}.

\begin{table}[htbp]
\caption{Dataset and Task Settings}
\centering
\setlength{\tabcolsep}{2pt}
\renewcommand{\arraystretch}{1}  
\fontsize{8}{9}\selectfont
\begin{tabular}{ccc}
\hline
\textbf{Datasets} & \textbf{Main Task} & \textbf{Target Property}  \\ \hline
\multirow{3}{*}{CIFAR-10} &\multirow{3}{*}{Image Classification}
 & Gray  \\
 && Speckle Noise  \\
 && Gaussian Noise  \\ \hline
\multirow{3}{*}{VGGFace2S}&\multirow{3}{*}{Face Classification} 
 & White Male  \\
 && Young Male  \\ 
 &&Black-Haired Female \\ \hline
\multirow{2}{*}{UTKFace}&\multirow{2}{*}{Race Classification} 
 & Child \\
 && Senior \\ \hline
\multirow{2}{*}{BreakHis}&\multirow{2}{*}{Benign vs. Malignant} 
 & Adenosis \\
 && Fibroadenoma \\ \hline
\end{tabular}
\label{tab:dataset_detail}
\end{table}

%\lnote{the detail of testset and trainset}
CIFAR-10 contains 50,000 training and 10,000 test samples for image classification. Since it lacks property annotations, we follow~\cite{wang2020towards,arora2024hybrid} and construct three synthetic properties by applying grayscale conversion, speckle noise, and Gaussian noise each to a randomly chosen 10\% subset, treating each transformation type as a distinct target property. 

VGGFace2S is constructed following \cite{tian2023manipulating}, by selecting the 50 subjects with the largest number of samples from VGGFace2 (which has over three million images of 9,131 subjects). Property annotations are from MAADFace~\cite{terhorst2021maad, terhorst2019reliable}, which provides rich demographic and appearance attribute labels. We select three target properties to match the experimental scale of CIFAR-10 and to cover diverse attribute types. %The training set contains over 20,000 images, with 100 images per subject reserved for testing. 
The training set contains over 20,000 images (23,754 for white male inference, 22,649 for young male, and 20,357 for black-haired female), with 100 images per subject reserved for testing. The variation in size is because we rely on the property annotations provided by MAADFace~\cite{terhorst2021maad}, which are not available for all samples in VGGFace2S. As a result, for each target property, only samples with the corresponding annotation are kept, leading to different training set sizes per property.

UTKFace contains over 23,000 facial images annotated with age, gender, and ethnicity, which are split into 80\% for training and 20\% for testing. BreakHis contains 7,909 microscopic breast cancer images, split into 65\% for training and 35\% for testing. For these two datasets, due to the limited annotation space and computational budget, we select two target properties each from the available annotations: child and senior for UTKFace, and adenosis and fibroadenoma for BreakHis. 

In all settings, the target property corresponds to a minority subgroup comprising less than 20\% of the dataset. This reflects a challenging yet realistic inference scenario, as inferring a property that holds for the majority of samples could be trivially achieved by random guessing. %More details about datasets are in Appendix~\ref{sec:dataset_detail}.

\textbf{Model Architectures.} We select three classical backbone architectures: 
VGG-16~\cite{simonyan2014very}, ResNet-56~\cite{he2016deep}, and 
MobileNet~\cite{howard2017mobilenets}. 
%For multi-exit network design, we follow the scheme proposed in~\cite{kaya2019shallow}, which has been extensively studied in the existing literature~\cite{dong2023mind,chen2023dark,li2022auditing,hong2020panda}.
Multi-exit networks are constructed following~\cite{kaya2019shallow, dong2023mind,chen2023dark,li2022auditing,hong2020panda}. For each experimental setting, similarly to \cite{li2022auditing,kaya2019shallow}, we randomly sample the number of exits from $[4, 6]$ and insert them at randomly chosen positions. 

%We conduct five independent trials for each setting and report the average and standard deviation of the results. By default, the last internal exit (immediately preceding the final exit) is used as the designated exit, and the confidence threshold for early exiting is uniformly set to 0.9. The parameter $\epsilon$ in Section~\ref{sec:loss-design-r1} is set to 0.6, and $\lambda$ in Equation~\ref{eq:target_loss} and Equation~\ref{eq:before_loss} is set to 1.5. 

\textbf{Hyperparameter Settings.} While our design is robust across different hyperparameter configurations (see hyperparameter studies in Section~\ref{sec:hyperparameter}), we optimize hyperparameters for each model architecture to ensure a fair comparison. The optimal settings are used in the paper. For all architectures, we use the exit immediately preceding the final exit as the designated exit, and set the confidence threshold for early termination to 0.9. The values of $\epsilon$ (Section~\ref{sec:loss-design-r1}) and $\lambda$ (Equation~\ref{eq:target_loss} and Equation~\ref{eq:before_loss}) are different for each architecture. $\epsilon = 0.6$ and $\lambda = 1.5$ for VGG-16; $\epsilon = 0.3$ and $\lambda = 0.5$ for ResNet-56; and  $\epsilon = 0.6$ and $\lambda = 0.5$ for MobileNet. The exception is BreakHis, where we set $\epsilon = 0.4$ because BreakHis involves binary classification, and large label smoothing $\epsilon$ can interfere with training and reduce main task accuracy.

\textbf{Attacker Training.} As described in Section~\ref{sec:loss-design-r3}, our training pipeline consists of two main steps: clean model training and fine-tuning with adversarial loss. For clean model training, following~\cite{kaya2019shallow}, we first train the backbone model and then insert internal exits (classifiers) based on the trained backbone. The backbone is trained using the SGD optimizer for 100 epochs, with an initial learning rate of 0.01 that is reduced by a factor of 10 at epochs 35, 65, and 85. For inserting the internal exits, we use the Adam optimizer for 25 epochs with a learning rate of 0.001, which is decayed by a factor of 10 at epoch 15. The batch size is set to 64 for all experiments. As for fine-tuning with adversarial loss , the clean multi-exit model is fine-tuned using the SGD optimizer for 100 epochs with a learning rate of 0.001.  We conduct five independent trials for each setting and report the average and standard deviation of the results.

% \textbf{Data Augmentation.} 
% We also implement a customized data augmentation strategy to facilitate attack model training. The proportion of samples with the target property in the training set is less than 20\% for all datasets.  However, optimizing the loss defined in Equation~\ref{eq:overall_loss} requires a sufficient number of samples with the target property. To address this, we design a strategy to increase the proportion of targeted samples. 

% Specifically, for each training batch, we mix targeted samples with non-targeted samples to create additional training samples:
% $$x = \beta x_w + (1 - \beta) x_{wo}$$
% where $\beta$ is randomly chosen from $[0.9, 1]$, $x_w$ denotes a sample with the target property, and $x_{wo}$ denotes a sample without the target property. The newly created samples are treated as having the target property, and both their class label $y$ and property label $z$ are assigned the same values as those of $x_w$ (i.e., no interpolation is applied to the labels).

\textbf{Input Privacy Inference.} The inference consists of two steps. First, the CSP measures the inference latency of each query and infers the corresponding exit index. Second, the CSP classifies inputs exiting at the designated exit $T$ as having the target property and all others as non-targeted. Since existing work~\cite{li2022auditing,11224001} and our results in Section~\ref{sec:latency}  demonstrate that exit indices can be reliably inferred from latency with accuracy exceeding 99\%, this work primarily focuses on designing the model manipulation framework for the second step. Therefore, the main results are reported under the assumption that exit indices are already known. Nevertheless, we also provide end-to-end evaluation results in Table~\ref{tab:main_results} and Table~\ref{tab:backbone} (rows labeled ``Our-E2E''), where exit indices are inferred from real latency measurements on a dedicated server equipped with an NVIDIA RTX 4090 GPU. We also conduct end-to-end experiments within a trusted execution environment (Section~\ref{sec:latency}). Due to the larger latency differences between exits in TEE environments, exit indices can be inferred more reliably, and the results are nearly identical to those obtained when exit indices are assumed to be known. Therefore, we report the GPU-based end-to-end results in the main paper, as GPU inference represents a more challenging scenario with greater uncertainty in latency-based exit index inference.

\subsection{Metrics} \label{sec:metrics}
We evaluate the proposed attack using two categories of metrics: effectiveness and stealthiness.

\textbf{Effectiveness Metrics.} As outlined in Section~\ref{sec:design-overview}, our attack aims to achieve both high precision (R1) and high recall (R2) when determining whether a given input has the target property. Accordingly, we adopt \textbf{Precision}, \textbf{Recall}, and \textbf{F1 score} (the harmonic mean of the first two) as effectiveness metrics. Precision is defined as the ratio of samples that truly possess the target property and are output at the designated exit $T$, to all samples output at that exit.
Recall is the ratio of samples with the target property that exit at $T$ to all samples with the target property. 

%We note that our evaluation assumes the exit index is obtainable without inferring it from latency in practice, as prior work has shown that such inference can be reliably achieved (with an accuracy exceeding 98\%)~\cite{li2022auditing, 11224001}.

To satisfy R3, the manipulated model must behave comparably to a clean deployment. 
We evaluate stealthiness by comparing the differences in \textbf{Model Accuracy} and \textbf{Inference Speedup} between the manipulated and clean models. Model accuracy is the main task performance of the deployed multi-exit model.
Inference speedup is defined as the average number of floating-point operations required for per-sample inference in the full backbone model, divided by that in the multi-exit model. 
By comparing these differences, we can measure the extent to which the attack interferes with the model's main task performance and inference acceleration. %Ideally, these differences should be minimal.

%To avoid user suspicion, the attack is designed to produce models that behave similarly to clean ones (R3 in Section~\ref{sec:design-overview}). We evaluate stealthiness by comparing the differences in  \textbf{Model Accuracy} and \textbf{Inference Speedup}  between manipulated models and clean ones. Model accuracy is the main task performance of the deployed multi-exit model. Inference speedup is defined as the average number of floating-point operations required for per-sample inference in the backbone model, divided by that in the multi-exit model. By comparing these differences, we can measure the extent to which the attack interferes with the model's main task performance and inference acceleration. Ideally, these differences should be minimal.

\begin{table*}[!htb]
\caption{Attack effectiveness and stealthiness across datasets and target properties using VGG-16. Effectiveness is measured by precision, recall, and F1 score; stealthiness by model accuracy and inference speedup. Bold indicates the best results.}
\centering
\renewcommand{\arraystretch}{1}  % 增加行高
\fontsize{8}{7}\selectfont
% \small
% \begin{tabular}{ll|cc|cc|cc|cc}
\begin{tabular}{cccccccc}
\toprule

 \textbf{Dataset}
& \textbf{Target Property}
& \textbf{Method}
& \textbf{Precision (\%)}
& \textbf{Recall (\%)}
& \textbf{F1 ($10^{-2}$)}
& \textbf{Accuracy (\%)}
& \textbf{Speedup ($\times$)}\\

\midrule
% cifar10-G
\multirow{15}{*}{\textbf{CIFAR-10}}
&\multirow{5}{*}{\textbf{Gray}}

& \textbf{RG} & 10.00\scalebox{0.8}{$\pm$0.00}&50.00\scalebox{0.8}{$\pm$0.00}&16.67\scalebox{0.8}{$\pm$0.00}&\textbf{91.27}\scalebox{0.8}{$\pm$0.23}&\textbf{2.16}\scalebox{0.8}{$\pm$0.07} \\

&& \textbf{FP} & 16.53\scalebox{0.8}{$\pm$0.85}&9.66\scalebox{0.8}{$\pm$3.19}&11.89\scalebox{0.8}{$\pm$2.44}&\textbf{91.27}\scalebox{0.8}{$\pm$0.23}&\textbf{2.16}\scalebox{0.8}{$\pm$0.07} \\

&& \textbf{LR} & 25.75\scalebox{0.8}{$\pm$6.39}&53.06\scalebox{0.8}{$\pm$9.73}&33.33\scalebox{0.8}{$\pm$4.62}&\textbf{91.27}\scalebox{0.8}{$\pm$0.23}&\textbf{2.16}\scalebox{0.8}{$\pm$0.07} \\

&& \textbf{Our}  & \textbf{77.48}\scalebox{0.8}{$\pm$9.25} & \textbf{84.98}\scalebox{0.8}{$\pm$2.88} &\textbf{80.71}\scalebox{0.8}{$\pm$5.44}&  91.06\scalebox{0.8}{$\pm$0.30}&2.08 \scalebox{0.8}{$\pm$0.08}\\
&& \textbf{Our-E2E}  & \textbf{77.46}\scalebox{0.8}{$\pm$9.26} & \textbf{84.82}\scalebox{0.8}{$\pm$2.75} &\textbf{80.63}\scalebox{0.8}{$\pm$5.43}&  91.06\scalebox{0.8}{$\pm$0.30}&2.08 \scalebox{0.8}{$\pm$0.08}\\
\cmidrule(lr){2-8}
% cifar10-SN

&\multirow{5}{*}{\textbf{Speckle Noise}}
& \textbf{RG} & 10.00\scalebox{0.8}{$\pm$0.00}&50.00\scalebox{0.8}{$\pm$0.00}&16.67\scalebox{0.8}{$\pm$0.00}& \textbf{90.97}\scalebox{0.8}{$\pm$0.21}&\textbf{2.10}\scalebox{0.8}{$\pm$0.08} \\

&& \textbf{FP} &21.66\scalebox{0.8}{$\pm$1.13}&15.36\scalebox{0.8}{$\pm$5.24}&17.54\scalebox{0.8}{$\pm$4.28}& \textbf{90.97}\scalebox{0.8}{$\pm$0.21}&\textbf{2.10}\scalebox{0.8}{$\pm$0.08} \\

&& \textbf{LR} & 15.70\scalebox{0.8}{$\pm$1.66}&53.76\scalebox{0.8}{$\pm$10.48}&23.95\scalebox{0.8}{$\pm$1.2}&\textbf{90.97}\scalebox{0.8}{$\pm$0.21}&\textbf{2.10}\scalebox{0.8}{$\pm$0.08} \\

&& \textbf{Our}  & \textbf{77.91}\scalebox{0.8}{$\pm$16.28} & \textbf{80.26}\scalebox{0.8}{$\pm$3.52} &\textbf{78.44}\scalebox{0.8}{$\pm$9.98}& 90.32\scalebox{0.8}{$\pm$0.40}& 2.07\scalebox{0.8}{$\pm$0.17}\\

&& \textbf{Our-E2E}  & \textbf{77.46}\scalebox{0.8}{$\pm$15.85} & \textbf{79.24}\scalebox{0.8}{$\pm$3.44} &\textbf{77.81}\scalebox{0.8}{$\pm$9.61}& 90.32\scalebox{0.8}{$\pm$0.40}& 2.07\scalebox{0.8}{$\pm$0.17}\\
\cmidrule(lr){2-8}

% cifar10_GN

&\multirow{5}{*}{\textbf{Gaussian Noise}}
& \textbf{RG} & 10.00\scalebox{0.8}{$\pm$0.00}&50.00\scalebox{0.8}{$\pm$0.00}&16.67\scalebox{0.8}{$\pm$0.00}&\textbf{90.72}\scalebox{0.8}{$\pm$0.14}&\textbf{2.05}\scalebox{0.8}{$\pm$0.08} \\

&& \textbf{FP} &24.31\scalebox{0.8}{$\pm$1.38}&19.16\scalebox{0.8}{$\pm$1.35}&21.40\scalebox{0.8}{$\pm$0.97}&\textbf{90.72}\scalebox{0.8}{$\pm$0.14}&\textbf{2.05}\scalebox{0.8}{$\pm$0.08} \\

&& \textbf{LR} & 16.59\scalebox{0.8}{$\pm$0.65}&54.10\scalebox{0.8}{$\pm$8.67}&25.22\scalebox{0.8}{$\pm$0.67}&\textbf{90.72}\scalebox{0.8}{$\pm$0.14}&\textbf{2.05}\scalebox{0.8}{$\pm$0.08} \\

&& \textbf{Our} &\textbf{78.14}\scalebox{0.8}{$\pm$10.80}&\textbf{81.22}\scalebox{0.8}{$\pm$4.30}&\textbf{79.12}\scalebox{0.8}{$\pm$5.11}& 90.40\scalebox{0.8}{$\pm$0.28}&2.02\scalebox{0.8}{$\pm$0.20}\\

&& \textbf{Our-E2E} &\textbf{77.64}\scalebox{0.8}{$\pm$10.78}&\textbf{79.64}\scalebox{0.8}{$\pm$3.44}&\textbf{78.09}\scalebox{0.8}{$\pm$4.60}& 90.40\scalebox{0.8}{$\pm$0.28}&2.02\scalebox{0.8}{$\pm$0.20}\\
\midrule

%vggmy
\multirow{15}{*}{\textbf{VGGFace2S}}
&\multirow{5}{*}{\textbf{Young Male}}

& \textbf{RG} &13.60\scalebox{0.8}{$\pm$0.00}&50.00\scalebox{0.8}{$\pm$0.00}&21.38\scalebox{0.8}{$\pm$0.00}&91.07\scalebox{0.8}{$\pm$0.41}&1.56\scalebox{0.8}{$\pm$0.04} \\

&& \textbf{FP} &20.54\scalebox{0.8}{$\pm$1.27}&14.85\scalebox{0.8}{$\pm$11.06}&15.62\scalebox{0.8}{$\pm$5.70}&91.07\scalebox{0.8}{$\pm$0.41}&1.56\scalebox{0.8}{$\pm$0.04} \\

&& \textbf{LR} & 17.76\scalebox{0.8}{$\pm$1.14}&42.68\scalebox{0.8}{$\pm$12.04}&24.41\scalebox{0.8}{$\pm$1.84}&91.07\scalebox{0.8}{$\pm$0.41}&1.56\scalebox{0.8}{$\pm$0.04} \\

&& \textbf{Our} &\textbf{93.19}\scalebox{1.05}{$\pm$0.43}&\textbf{80.71}\scalebox{0.8}{$\pm$2.77}&\textbf{86.46}\scalebox{0.8}{$\pm$1.24}&\textbf{93.14}\scalebox{0.8}{$\pm$0.43}&\textbf{1.67}\scalebox{0.8}{$\pm$0.04}\\
&& \textbf{Our-E2E} &\textbf{92.86}\scalebox{1.05}{$\pm$0.94}&\textbf{80.68}\scalebox{0.8}{$\pm$2.73}&\textbf{86.29}\scalebox{0.8}{$\pm$1.25}&\textbf{93.14}\scalebox{0.8}{$\pm$0.43}&\textbf{1.67}\scalebox{0.8}{$\pm$0.04}\\
\cmidrule(lr){2-8}

%vggmw

&\multirow{5}{*}{\textbf{White Male}}

& \textbf{RG} &14.50\scalebox{0.8}{$\pm$0.00}&50.00\scalebox{0.8}{$\pm$0.00}&22.48\scalebox{0.8}{$\pm$0.00}&91.80\scalebox{0.8}{$\pm$0.27}&1.51\scalebox{0.8}{$\pm$0.05}\\
&& \textbf{FP} &36.45\scalebox{0.8}{$\pm$4.83}&15.86\scalebox{0.8}{$\pm$6.04}&21.26\scalebox{0.8}{$\pm$6.58}&91.80\scalebox{0.8}{$\pm$0.27}&1.51\scalebox{0.8}{$\pm$0.05}\\
&& \textbf{LR} &20.32\scalebox{0.8}{$\pm$0.88}&50.40\scalebox{0.8}{$\pm$12.63}&28.44\scalebox{0.8}{$\pm$3.02}&91.80\scalebox{0.8}{$\pm$0.27}&1.51\scalebox{0.8}{$\pm$0.05}\\

&& \textbf{Our} &\textbf{96.37}\scalebox{0.8}{$\pm$0.67}&\textbf{79.48}\scalebox{0.8}{$\pm$2.66}&\textbf{87.08}\scalebox{0.8}{$\pm$1.45}&\textbf{93.94}\scalebox{0.8}{$\pm$0.37}&\textbf{1.57}\scalebox{0.8}{$\pm$0.03}\\
&& \textbf{Our-E2E} &\textbf{95.81}\scalebox{0.8}{$\pm$1.20}&\textbf{79.34}\scalebox{0.8}{$\pm$2.69}&\textbf{86.76}\scalebox{0.8}{$\pm$1.36}&\textbf{93.94}\scalebox{0.8}{$\pm$0.37}&\textbf{1.57}\scalebox{0.8}{$\pm$0.03}\\
\cmidrule(lr){2-8}

%vggmbh

&\multirow{5}{*}{\textbf{\shortstack[c]{Black-Haired \\ Female}}}

& \textbf{RG} &13.02\scalebox{0.8}{$\pm$0.00}&50.00\scalebox{0.8}{$\pm$0.00}&20.66\scalebox{0.8}{$\pm$0.00}&89.56\scalebox{0.8}{$\pm$0.25}&1.44\scalebox{0.8}{$\pm$0.04} \\

&& \textbf{FP} &20.06\scalebox{0.8}{$\pm$2.27}&21.84\scalebox{0.8}{$\pm$7.36}&20.22\scalebox{0.8}{$\pm$2.94}&89.56\scalebox{0.8}{$\pm$0.25}&1.44\scalebox{0.8}{$\pm$0.04} \\

&& \textbf{LR} & 15.52\scalebox{0.8}{$\pm$0.74}&55.82\scalebox{0.8}{$\pm$18.20}&23.78\scalebox{0.8}{$\pm$2.47}&89.56\scalebox{0.8}{$\pm$0.25}&1.44\scalebox{0.8}{$\pm$0.04} \\

&& \textbf{Our} &\textbf{90.35}\scalebox{0.8}{$\pm$1.48}&\textbf{60.28}\scalebox{0.8}{$\pm$5.16}&\textbf{72.15}\scalebox{0.8}{$\pm$3.53}&\textbf{91.42}\scalebox{0.8}{$\pm$0.58}&\textbf{1.52}\scalebox{0.8}{$\pm$0.03}\\
&& \textbf{Our-E2E} &\textbf{90.02}\scalebox{0.8}{$\pm$1.52}&\textbf{60.19}\scalebox{0.8}{$\pm$5.15}&\textbf{71.99}\scalebox{0.8}{$\pm$3.60}&\textbf{91.42}\scalebox{0.8}{$\pm$0.58}&\textbf{1.52}\scalebox{0.8}{$\pm$0.03}\\
\midrule

%UTKC
\multirow{10}{*}{\textbf{UTKFace}}
&\multirow{5}{*}{\textbf{Child}}

& \textbf{RG} &16.41\scalebox{0.8}{$\pm$0.00}&50.00\scalebox{0.8}{$\pm$0.00}&24.71\scalebox{0.8}{$\pm$0.00}&83.81\scalebox{0.8}{$\pm$0.29}&1.50\scalebox{0.8}{$\pm$0.06} \\

&& \textbf{FP}&18.53\scalebox{0.8}{$\pm$0.33}&16.07\scalebox{0.8}{$\pm$4.30}&16.94\scalebox{0.8}{$\pm$2.66} &83.81\scalebox{0.8}{$\pm$0.29}&1.50\scalebox{0.8}{$\pm$0.06} \\

&& \textbf{LR} & 17.13\scalebox{0.8}{$\pm$0.50}&49.51\scalebox{0.8}{$\pm$10.05}&25.21\scalebox{0.8}{$\pm$1.59}&83.81\scalebox{0.8}{$\pm$0.29}&1.50\scalebox{0.8}{$\pm$0.06} \\

&& \textbf{Our} &\textbf{97.33}\scalebox{0.8}{$\pm$0.38}&\textbf{61.10}\scalebox{0.8}{$\pm$3.96}&\textbf{75.00}\scalebox{0.8}{$\pm$3.02}&\textbf{85.25}\scalebox{0.8}{$\pm$0.21}&\textbf{1.55}\scalebox{0.8}{$\pm$0.04} \\
&& \textbf{Our-E2E} &\textbf{95.89}\scalebox{0.8}{$\pm$2.45}&\textbf{61.10}\scalebox{0.8}{$\pm$3.96}&\textbf{74.52}\scalebox{0.8}{$\pm$2.44}&\textbf{85.25}\scalebox{0.8}{$\pm$0.21}&\textbf{1.55}\scalebox{0.8}{$\pm$0.04} \\
\cmidrule(lr){2-8}

%UTKS

&\multirow{5}{*}{\textbf{Senior}}

& \textbf{RG} &19.34\scalebox{0.8}{$\pm$0.00}&50.00\scalebox{0.8}{$\pm$0.00}&27.89\scalebox{0.8}{$\pm$0.00}&84.28\scalebox{0.8}{$\pm$0.47}&\textbf{1.47}\scalebox{0.8}{$\pm$0.06} \\

&& \textbf{FP} &28.79\scalebox{0.8}{$\pm$3.45}&27.46\scalebox{0.8}{$\pm$5.99}&27.44\scalebox{0.8}{$\pm$2.53}&84.28\scalebox{0.8}{$\pm$0.47}&\textbf{1.47}\scalebox{0.8}{$\pm$0.06} \\

&& \textbf{LR} & 26.18\scalebox{0.8}{$\pm$1.33}&39.15\scalebox{0.8}{$\pm$8.67}&30.79\scalebox{0.8}{$\pm$2.18}&84.28\scalebox{0.8}{$\pm$0.47}&\textbf{1.47}\scalebox{0.8}{$\pm$0.06} \\

&& \textbf{Our}  &\textbf{89.36}\scalebox{0.8}{$\pm$1.02} &\textbf{50.20}\scalebox{0.8}{$\pm$1.58} &\textbf{64.27}\scalebox{0.8}{$\pm$1.20}&\textbf{85.22}\scalebox{0.8}{$\pm$0.15}&1.43\scalebox{0.8}{$\pm$0.04}\\

&& \textbf{Our-E2E}  &\textbf{89.32}\scalebox{0.8}{$\pm$1.07} &\textbf{50.16}\scalebox{0.8}{$\pm$1.59} &\textbf{64.22}\scalebox{0.8}{$\pm$1.21}&\textbf{85.22}\scalebox{0.8}{$\pm$0.15}&1.43\scalebox{0.8}{$\pm$0.04}\\
\midrule

%BreakA
\multirow{10}{*}{\textbf{BreakHis}}
&\multirow{5}{*}{\textbf{Adenosis}}

& \textbf{RG} &6.58\scalebox{0.8}{$\pm$0.00}&50.00\scalebox{0.8}{$\pm$0.00}&11.63\scalebox{0.8}{$\pm$0.00}&\textbf{85.80}\scalebox{0.8}{$\pm$0.54}&\textbf{3.17}\scalebox{0.8}{$\pm$0.28} \\

&& \textbf{FP} &20.88\scalebox{0.8}{$\pm$4.87}&9.63\scalebox{0.8}{$\pm$6.40}&11.84\scalebox{0.8}{$\pm$6.36}&\textbf{85.80}\scalebox{0.8}{$\pm$0.54}&\textbf{3.17}\scalebox{0.8}{$\pm$0.28} \\

&& \textbf{LR} & 12.21\scalebox{0.8}{$\pm$5.77}&34.45\scalebox{0.8}{$\pm$21.37}&13.7\scalebox{0.8}{$\pm$3.33}&\textbf{85.80}\scalebox{0.8}{$\pm$0.54}&\textbf{3.17}\scalebox{0.8}{$\pm$0.28} \\

&& \textbf{Our} &\textbf{43.08}\scalebox{0.8}{$\pm$6.75}&\textbf{53.61}\scalebox{0.8}{$\pm$8.05}&\textbf{46.93}\scalebox{0.8}{$\pm$3.46}&84.55\scalebox{0.8}{$\pm$0.87}&2.62\scalebox{0.8}{$\pm$0.07} \\
&& \textbf{Our-E2E} &\textbf{42.40}\scalebox{0.8}{$\pm$6.72}&\textbf{53.09}\scalebox{0.8}{$\pm$8.62}&\textbf{46.24}\scalebox{0.8}{$\pm$3.50}&84.55\scalebox{0.8}{$\pm$0.87}&2.62\scalebox{0.8}{$\pm$0.07} \\
\cmidrule(lr){2-8}

% BreakKF

&\multirow{5}{*}{\textbf{Fibroadenoma}}

& \textbf{RG} &11.40\scalebox{0.8}{$\pm$0.00}&50.00\scalebox{0.8}{$\pm$0.00}&18.57\scalebox{0.8}{$\pm$0.00}&\textbf{86.11}\scalebox{0.8}{$\pm$0.54}&\textbf{2.89}\scalebox{0.8}{$\pm$0.45} \\
&& \textbf{FP} &16.82\scalebox{0.8}{$\pm$3.56}&35.71\scalebox{0.8}{$\pm$37.71}&16.38\scalebox{0.8}{$\pm$9.67}&\textbf{86.11}\scalebox{0.8}{$\pm$0.54}&\textbf{2.89}\scalebox{0.8}{$\pm$0.45} \\
&& \textbf{LR} & 9.89\scalebox{0.8}{$\pm$4.07}&23.93\scalebox{0.8}{$\pm$16.58}&13.72\scalebox{0.8}{$\pm$6.65}&\textbf{86.11}\scalebox{0.8}{$\pm$0.54}&\textbf{2.89}\scalebox{0.8}{$\pm$0.45} \\
&& \textbf{Our}&\textbf{45.22}\scalebox{0.8}{$\pm$0.90}&\textbf{78.19}\scalebox{0.8}{$\pm$2.43}&\textbf{ 57.27}\scalebox{0.8}{$\pm$0.39} &85.52\scalebox{0.8}{$\pm$0.72}&2.18\scalebox{0.8}{$\pm$0.05} \\
&& \textbf{Our-E2E}&\textbf{44.91}\scalebox{0.8}{$\pm$0.94}&\textbf{77.95}\scalebox{0.8}{$\pm$2.78}&\textbf{ 56.95}\scalebox{0.8}{$\pm$0.67} &85.52\scalebox{0.8}{$\pm$0.72}&2.18\scalebox{0.8}{$\pm$0.05} \\
\bottomrule
\end{tabular}

\label{tab:main_results}
\end{table*}

\subsection{Baselines} \label{sec:baselines}
As discussed in Section~\ref{sec:multi-exit-leakage}, no existing work studies input property inference through active manipulation of multi-exit models. We therefore construct three baselines.

The first baseline is \textbf{Random Guessing (RG)}, where the attacker predicts the 
presence of the target property with a fixed probability of 50\% for each input. The second baseline is \textbf{Frequency based Prediction (FP)}, which is adapted from the passive method proposed by Kannan et al.~\cite{kannan2023prediction}. The original method predicts the output class of input samples, by analyzing the output pattern at each exit and predicting that all samples exiting at a given exit belong to the most frequently occurring class at that exit.
We adapt this method to our setting as follows. We compute the proportion of targeted inputs that exit at each exit, and designate the exit with the highest proportion as the indicator exit. We then predict that all samples exiting at that indicator exit carry the target property, while samples exiting at all other exits are predicted as non-targeted.  The third baseline is \textbf{Logistic Regression Classifier (LR)}, which treats exit-routing information as features and learns a binary classifier to predict whether an input possesses the target property. Specifically, we first record the exit indices of training samples through the trained multi-exit model. We then train a logistic regression model using the exit indices as input features and the corresponding property labels as targets.

\subsection{Main Results} \label{sec:main_results}

We evaluate the proposed attack on four datasets across ten target properties (Table~\ref{tab:dataset_detail}) using VGG-16 as the backbone. Results for other backbone architectures show similar trends and are in Section~\ref{sec:other_backbone}. For each setting, we repeat the experiment five times and report the average and standard deviation.

\textbf{Experimental Results.}
Table~\ref{tab:main_results} presents the results. The ``RG'', ``FP'', and ``LR'' rows correspond to the three baselines introduced in Section~\ref{sec:baselines}. The ``Our'' and ``Our-E2E'' rows represent the results of our method under the setting where exit indices are known and the end-to-end GPU evaluation setting with real latency, respectively. Our attack consistently and significantly outperforms all baselines across all settings while maintaining stealthiness. 

\textit{Attack effectiveness.} The attack precision is improved by a large margin. Baseline methods achieve precision below 25\% in seven out of ten settings and below $36.5\%$ in all settings. In contrast, our attack achieves precision above $90\%$ in four settings, exceeding $77\%$ in eight settings, and above $42\%$ in all settings. Although precision on the BreakHis dataset is relatively lower (still above $42\%$), it shows a substantial improvement over the baselines, which fall below $21\%$. Recall is also consistently improved across all settings. 
The best recall achieved among the three baselines averages 51.71\% across all settings.
%The average of the best recall values of the three baselines is 51.71\%. 
Our method improves the average recall to 71\% (70.62\% under the end-to-end setting). F1 scores show the same trend. Our method achieves F1 scores above 0.71 in seven out of ten settings and above 0.46 in all settings, while baseline F1 scores are less than 0.34 across all settings.

\textit{Attack stealthiness.} Our attack largely preserves task accuracy and inference speedup. Main task accuracy drops by at most 1.3\% relative to clean models across all settings. In some settings, main task accuracy even slightly improves, which can be attributed to our training pipeline being explicitly designed to preserve model performance. 
Inference speedup is similarly preserved, with decrease below 0.1 in most settings. The exceptions are the two properties on BreakHis, where inference speedup decreases by 0.55 and 0.71. We attribute this to the nature of the target property. Unlike the target properties on other datasets, which are orthogonal to or simpler than the main task, the target property on BreakHis involves inferring histopathological subtypes, which is a substantially more complex task than the model's main task of binary cancer detection. Introducing this additional objective during fine-tuning increases the demand on the model's capacity, which disrupts the confidence distributions at intermediate exits and causes more samples to propagate to deeper layers, reducing average speedup. Nonetheless, the manipulated models still achieve speedups of 2.62$\times$ and 2.18$\times$, while accuracy drops by less than 1.3\%. Thus, the attack still preserves task accuracy and substantial inference benefits in these cases.

\textit{Known exit indices versus end-to-end inference.} The results under the known-exit-index setting (the ``Our'' rows) and the end-to-end setting (the ``Our-E2E'' rows) are nearly identical. Differences in attack effectiveness are less than 0.5\% in 21 out of 30 cases (across all datasets, properties, and metrics) and less than 1.58\% in all cases. These results confirm that our attack remains effective under realistic deployment conditions and is robust to latency noise.

In summary, our attack achieves strong effectiveness while maintaining stealthiness across diverse datasets and target properties.

\subsection{Results on Different Backbones} \label{sec:other_backbone}
To evaluate the effectiveness of our attack across different model architectures, we select VGG-16, ResNet-56, and MobileNet as backbones, and conduct experiments on CIFAR-10 and VGGFace2S with gray and white male as the target property, respectively. 
%The results for VGG-16 are reused from Table~\ref{tab:main_results}. %For fair comparison, hyperparameters are optimized for each architecture (details in Appendix~\ref{sec:training_detail}).

\textbf{Experimental Results.} Table~\ref{tab:backbone} presents the results. Our attack consistently outperforms all three baselines across all settings. Attack precision exceeds 77\%, recall exceeds 67\%, and F1 scores exceed 0.73, while baseline precision, recall, and F1 remain below 41\%, 61.5\%, and 0.334, respectively. Main task accuracy degrades by less than 0.5\%, and inference speedup is also well preserved for VGG-16 and ResNet-56, with decrease below 0.1. MobileNet shows slightly larger reductions, from 2.79$\times$ to 2.04$\times$ on CIFAR-10 and from 1.75$\times$ to 1.45$\times$ on VGGFace2S. One possible explanation for the decrease is that MobileNet, as a lightweight architecture with fewer parameters, may need to allocate more of its limited capacity to accommodate the manipulation objective while preserving the original inference behavior. Nonetheless, the speedup remains substantial ($2.04\times$ on CIFAR-10), and the degradation alone does not necessarily indicate manipulation, as lightweight models naturally operate under tighter resource constraints and may exhibit larger variations in inference efficiency across deployment environments.

\begin{table*}[htb]
\caption{Attack effectiveness across backbone architectures on CIFAR-10 and VGGFace2S. Target properties are gray and white male, respectively. The results for VGG-16 are reused from Table~\ref{tab:main_results}}
\centering
\renewcommand{\arraystretch}{1}  % 增加行高
\fontsize{8}{7}\selectfont

\begin{tabular}{cccccccc}
\toprule

 \textbf{Backbone}
& \textbf{Dataset}
& \textbf{Method}
& \textbf{Precision (\%)}
& \textbf{Recall (\%)}
& \textbf{F1 ($10^{-2}$)}
& \textbf{Accuracy (\%)}
& \textbf{Speedup ($\times$)}\\

\midrule

% VGG
\multirow{10}{*}{\textbf{VGG-16}}
&\multirow{5}{*}{\textbf{CIFAR-10}}
& \textbf{RG} & 10.00\scalebox{0.8}{$\pm$0.00}&50.00\scalebox{0.8}{$\pm$0.00}&16.67\scalebox{0.8}{$\pm$0.00}&\textbf{91.27}\scalebox{0.8}{$\pm$0.23}&\textbf{2.16}\scalebox{0.8}{$\pm$0.07} \\

&& \textbf{FP} & 16.53\scalebox{0.8}{$\pm$0.85}&9.66\scalebox{0.8}{$\pm$3.19}&11.89\scalebox{0.8}{$\pm$2.44}&\textbf{91.27}\scalebox{0.8}{$\pm$0.23}&\textbf{2.16}\scalebox{0.8}{$\pm$0.07} \\

&& \textbf{LR} & 25.75\scalebox{0.8}{$\pm$6.39}&53.06\scalebox{0.8}{$\pm$9.73}&33.33\scalebox{0.8}{$\pm$4.62}&\textbf{91.27}\scalebox{0.8}{$\pm$0.23}&\textbf{2.16}\scalebox{0.8}{$\pm$0.07} \\

&& \textbf{Our}  & \textbf{77.48}\scalebox{0.8}{$\pm$9.25} & \textbf{84.98}\scalebox{0.8}{$\pm$2.88} &\textbf{80.71}\scalebox{0.8}{$\pm$5.44}&  91.06\scalebox{0.8}{$\pm$0.30}&2.08 \scalebox{0.8}{$\pm$0.08}\\
&& \textbf{Our-E2E}  & \textbf{77.46}\scalebox{0.8}{$\pm$9.26} & \textbf{84.82}\scalebox{0.8}{$\pm$2.75} &\textbf{80.63}\scalebox{0.8}{$\pm$5.43}&  91.06\scalebox{0.8}{$\pm$0.30}&2.08 \scalebox{0.8}{$\pm$0.08}\\
\cmidrule(lr){2-8}

&\multirow{5}{*}{\textbf{VGGFace2S}}
& \textbf{RG} &14.50\scalebox{0.8}{$\pm$0.00}&50.00\scalebox{0.8}{$\pm$0.00}&22.48\scalebox{0.8}{$\pm$0.00}&91.80\scalebox{0.8}{$\pm$0.27}&1.51\scalebox{0.8}{$\pm$0.05}\\
&& \textbf{FP} &36.45\scalebox{0.8}{$\pm$4.83}&15.86\scalebox{0.8}{$\pm$6.04}&21.26\scalebox{0.8}{$\pm$6.58}&91.80\scalebox{0.8}{$\pm$0.27}&1.51\scalebox{0.8}{$\pm$0.05}\\
&& \textbf{LR} &20.32\scalebox{0.8}{$\pm$0.88}&50.40\scalebox{0.8}{$\pm$12.63}&28.44\scalebox{0.8}{$\pm$3.02}&91.80\scalebox{0.8}{$\pm$0.27}&1.51\scalebox{0.8}{$\pm$0.05}\\

&& \textbf{Our} &\textbf{96.37}\scalebox{0.8}{$\pm$0.67}&\textbf{79.48}\scalebox{0.8}{$\pm$2.66}&\textbf{87.08}\scalebox{0.8}{$\pm$1.45}&\textbf{93.94}\scalebox{0.8}{$\pm$0.37}&\textbf{1.57}\scalebox{0.8}{$\pm$0.03}\\
&& \textbf{Our-E2E} &\textbf{95.81}\scalebox{0.8}{$\pm$1.20}&\textbf{79.34}\scalebox{0.8}{$\pm$2.69}&\textbf{86.76}\scalebox{0.8}{$\pm$1.36}&\textbf{93.94}\scalebox{0.8}{$\pm$0.37}&\textbf{1.57}\scalebox{0.8}{$\pm$0.03}\\
\midrule

% resnet
\multirow{10}{*}{\textbf{ResNet-56}}
&\multirow{5}{*}{\textbf{CIFAR-10}}
&\multirow{1}{*}{\textbf{RG}}&10.00\scalebox{0.8}{$\pm$0.00}&50.00\scalebox{0.8}{$\pm$0.00}&16.67\scalebox{0.8}{$\pm$0.00}&\textbf{88.10}\scalebox{0.8}{$\pm$0.24}&1.57\scalebox{0.8}{$\pm$0.08}\\
&&\multirow{1}{*}{\textbf{FP}}&14.08\scalebox{0.8}{$\pm$0.85}&18.78\scalebox{0.8}{$\pm$7.70}&15.52\scalebox{0.8}{$\pm$2.14}&\textbf{88.10}\scalebox{0.8}{$\pm$0.24}&1.57\scalebox{0.8}{$\pm$0.08}\\
&&\multirow{1}{*}{\textbf{LR}}& 11.70\scalebox{0.8}{$\pm$0.01}&61.48\scalebox{0.8}{$\pm$6.58}&19.63\scalebox{0.8}{$\pm$0.35}&\textbf{88.10}\scalebox{0.8}{$\pm$0.24}&1.57\scalebox{0.8}{$\pm$0.08}\\
&&\multirow{1}{*}{\textbf{Our}}&\textbf{80.38}\scalebox{0.8}{$\pm$5.61}&\textbf{67.86}\scalebox{0.8}{$\pm$9.13}&\textbf{73.42}\scalebox{0.8}{$\pm$7.18}&87.67\scalebox{0.8}{$\pm$0.80}&\textbf{1.73}\scalebox{0.8}{$\pm$0.08}\\
&&\multirow{1}{*}{\textbf{Our-E2E}}&\textbf{79.83}\scalebox{0.8}{$\pm$5.27}&\textbf{67.55}\scalebox{0.8}{$\pm$8.03}&\textbf{73.03}\scalebox{0.8}{$\pm$6.52}&87.67\scalebox{0.8}{$\pm$0.80}&\textbf{1.73}\scalebox{0.8}{$\pm$0.08}\\
\cmidrule(lr){2-8}

&\multirow{5}{*}{\textbf{VGGFace2S}}
&\multirow{1}{*}{\textbf{RG}}&14.50\scalebox{0.8}{$\pm$0.00}&50.00\scalebox{0.8}{$\pm$0.00}&22.48\scalebox{0.8}{$\pm$0.00}&\textbf{86.19}\scalebox{0.8}{$\pm$0.51}&1.32\scalebox{0.8}{$\pm$0.04}\\
&&\multirow{1}{*}{\textbf{FP}}&40.70\scalebox{0.8}{$\pm$4.61}&5.16\scalebox{0.8}{$\pm$2.24}&8.98\scalebox{0.8}{$\pm$3.52}&\textbf{86.19}\scalebox{0.8}{$\pm$0.51}&1.32\scalebox{0.8}{$\pm$0.04}\\
&&\multirow{1}{*}{\textbf{LR}}& 22.42\scalebox{0.8}{$\pm$2.00}&43.20\scalebox{0.8}{$\pm$13.00}&29.04\scalebox{0.8}{$\pm$5.39}&\textbf{86.19}\scalebox{0.8}{$\pm$0.51}&1.32\scalebox{0.8}{$\pm$0.04}\\
&&\multirow{1}{*}{\textbf{Our}}&\textbf{90.40}\scalebox{0.8}{$\pm$1.13}&\textbf{73.05}\scalebox{0.8}{$\pm$2.50}&\textbf{80.78}\scalebox{0.8}{$\pm$1.38}&86.16\scalebox{0.8}{$\pm$0.56}&\textbf{1.36}\scalebox{0.8}{$\pm$0.04}\\
&&\multirow{1}{*}{\textbf{Our-E2E}}&\textbf{90.24}\scalebox{0.8}{$\pm$1.04}&\textbf{72.63}\scalebox{0.8}{$\pm$2.08}&\textbf{80.46}\scalebox{0.8}{$\pm$1.07}&86.16\scalebox{0.8}{$\pm$0.56}&\textbf{1.36}\scalebox{0.8}{$\pm$0.04}\\
\midrule

% mobilenet
\multirow{10}{*}{\textbf{MobileNet}}

&\multirow{5}{*}{\textbf{CIFAR-10}}
&\multirow{1}{*}{\textbf{RG}}&10.00\scalebox{0.8}{$\pm$0.00}&50.00\scalebox{0.8}{$\pm$0.00}&16.67\scalebox{0.8}{$\pm$0.00}&87.73\scalebox{0.8}{$\pm$0.26}&\textbf{2.79}\scalebox{0.8}{$\pm$0.07}\\
&&\multirow{1}{*}{\textbf{FP}}&15.63\scalebox{0.8}{$\pm$0.92}&9.60\scalebox{0.8}{$\pm$1.39}&11.85\scalebox{0.8}{$\pm$1.15}&87.73\scalebox{0.8}{$\pm$0.26}&\textbf{2.79}\scalebox{0.8}{$\pm$0.07}\\
&&\multirow{1}{*}{\textbf{LR}}& 12.30\scalebox{0.8}{$\pm$0.42}&57.68\scalebox{0.8}{$\pm$11.99}&20.11\scalebox{0.8}{$\pm$0.57}&87.73\scalebox{0.8}{$\pm$0.26}&\textbf{2.79}\scalebox{0.8}{$\pm$0.07}\\
&&\multirow{1}{*}{\textbf{Our}}&\textbf{83.08}\scalebox{0.8}{$\pm$6.76}&\textbf{69.24}\scalebox{0.8}{$\pm$4.90}&\textbf{75.21}\scalebox{0.8}{$\pm$1.69}& \textbf{87.84}\scalebox{0.8}{$\pm$0.90}&2.04\scalebox{0.8}{$\pm$0.12}\\
&&\multirow{1}{*}{\textbf{Our-E2E}}&\textbf{82.61}\scalebox{0.8}{$\pm$6.03}&\textbf{68.92}\scalebox{0.8}{$\pm$4.19}&\textbf{74.84}\scalebox{0.8}{$\pm$1.46}& \textbf{87.84}\scalebox{0.8}{$\pm$0.90}&2.04\scalebox{0.8}{$\pm$0.12}\\
\cmidrule(lr){2-8}
&\multirow{5}{*}{\textbf{VGGFace2S}}
&\multirow{1}{*}{\textbf{RG}}&14.50\scalebox{0.8}{$\pm$0.00}&50.00\scalebox{0.8}{$\pm$0.00}&22.48\scalebox{0.8}{$\pm$0.00}&88.64\scalebox{0.8}{$\pm$0.54}&\textbf{1.75}\scalebox{0.8}{$\pm$0.10}\\
&&\multirow{1}{*}{\textbf{FP}}&34.51\scalebox{0.8}{$\pm$3.28}&18.04\scalebox{0.8}{$\pm$9.56}&22.51\scalebox{0.8}{$\pm$6.93}& 88.64\scalebox{0.8}{$\pm$0.54}&\textbf{1.75}\scalebox{0.8}{$\pm$0.10}\\
&&\multirow{1}{*}{\textbf{LR}}& 22.41\scalebox{0.8}{$\pm$4.59}&54.04\scalebox{0.8}{$\pm$6.83}&30.95\scalebox{0.8}{$\pm$2.43}& 88.64\scalebox{0.8}{$\pm$0.54}&\textbf{1.75}\scalebox{0.8}{$\pm$0.10}\\
&&\multirow{1}{*}{\textbf{Our}}&\textbf{94.65}\scalebox{0.8}{$\pm$1.11}&\textbf{73.35}\scalebox{0.8}{$\pm$1.49}&\textbf{82.64}\scalebox{0.8}{$\pm$0.65} &\textbf{89.04}\scalebox{0.8}{$\pm$1.20}&1.45\scalebox{0.8}{$\pm$0.04}\\
&&\multirow{1}{*}{\textbf{Our-E2E}}&\textbf{94.24}\scalebox{0.8}{$\pm$1.26}&\textbf{73.08}\scalebox{0.8}{$\pm$1.40}&\textbf{82.30}\scalebox{0.8}{$\pm$0.52} &\textbf{89.04}\scalebox{0.8}{$\pm$1.20}&1.45\scalebox{0.8}{$\pm$0.04}\\
\bottomrule       
\end{tabular}

\label{tab:backbone}
\end{table*}

\subsection{Inferring Exit Index from Latency} \label{sec:latency}

\begin{table*}[htb]
\caption{Per-exit inference latency (ms) and exit inference accuracy (\%) across backbone architectures and deployment environments.}
\centering
\renewcommand{\arraystretch}{1}  % 增加行高
\fontsize{8}{7}\selectfont

\begin{tabular}{ccccccccc}
\toprule

 \textbf{Backbone}
& \textbf{Environment}
& \textbf{Exit 0}
& \textbf{Exit 1}
& \textbf{Exit 2}
& \textbf{Exit 3}
& \textbf{Exit 4}
& \textbf{Exit 5}
& \textbf{Inference Acc}\\

\midrule

% VGG
\multirow{2}{*}{\textbf{VGG-16}}&\textbf{TEE}&2.27\scalebox{0.8}{$\pm$0.01}&4.54\scalebox{0.8}{$\pm$0.01}&6.79\scalebox{0.8}{$\pm$0.02}&9.59\scalebox{0.8}{$\pm$0.03}&16.15\scalebox{0.8}{$\pm$0.04}&18.37\scalebox{0.8}{$\pm$0.05}&99.91\scalebox{0.8}{$\pm$0.06}\\
&\textbf{GPU}&0.41\scalebox{0.8}{$\pm$0.00}&0.73\scalebox{0.8}{$\pm$0.01}&1.02\scalebox{0.8}{$\pm$0.01}&1.33\scalebox{0.8}{$\pm$0.01}&1.68\scalebox{0.8}{$\pm$0.02}&1.96\scalebox{0.8}{$\pm$0.02}&99.62\scalebox{0.8}{$\pm$0.10}\\
\midrule

% VGG
\multirow{2}{*}{\textbf{ResNet-56}}&\textbf{TEE}&1.68\scalebox{0.8}{$\pm$0.00}&3.66\scalebox{0.8}{$\pm$0.01}&6.17\scalebox{0.8}{$\pm$0.02}&7.91\scalebox{0.8}{$\pm$0.02}&10.96\scalebox{0.8}{$\pm$0.02}&12.74\scalebox{0.8}{$\pm$0.02}&99.90\scalebox{0.8}{$\pm$0.11}\\
&\textbf{GPU}&0.67\scalebox{0.8}{$\pm$0.03}&1.32\scalebox{0.8}{$\pm$0.07}&2.08\scalebox{0.8}{$\pm$0.09}&2.70\scalebox{0.8}{$\pm$0.02}&3.72\scalebox{0.8}{$\pm$0.04}&4.40\scalebox{0.8}{$\pm$0.06}&99.20\scalebox{0.8}{$\pm$0.37}\\
\midrule

% VGG
\multirow{2}{*}{\textbf{MobileNet}}&\textbf{TEE}&2.05\scalebox{0.8}{$\pm$0.01}&3.18\scalebox{0.8}{$\pm$0.01}&4.45\scalebox{0.8}{$\pm$0.01}&5.72\scalebox{0.8}{$\pm$0.02}&6.83\scalebox{0.8}{$\pm$0.02}&7.87\scalebox{0.8}{$\pm$0.02}&99.81\scalebox{0.8}{$\pm$0.06}\\
&\textbf{GPU}&0.68\scalebox{0.8}{$\pm$0.01}&1.04\scalebox{0.8}{$\pm$0.01}&1.39\scalebox{0.8}{$\pm$0.02}&1.73\scalebox{0.8}{$\pm$0.02}&1.99\scalebox{0.8}{$\pm$0.02}&2.28\scalebox{0.8}{$\pm$0.02}&99.48\scalebox{0.8}{$\pm$0.41}\\

\bottomrule       
\end{tabular}

\label{tab:latencys}
\end{table*}

To evaluate the reliability of exit index inference from latency in real deployment environments, we conduct experiments on two hardware settings. The first is a \textbf{Trusted Execution Environment (TEE)}. We deploy the inference service on an Alibaba Cloud ECS server running Alibaba Cloud Linux 3, equipped with an Intel Xeon Platinum 8369B processor supporting SGX2 and 14 GB memory. We use Occlum 0.31.0 and run PyTorch models under hardware SGX. The Python runtime, PyTorch framework, model weights, and datasets are all encapsulated within the Enclave. The second is a \textbf{GPU Environment}. The platform is equipped with an Intel Core i9-13900KF processor, 128 GB memory, 4 TB storage, and an NVIDIA RTX 4090 GPU.

We evaluate three backbone architectures, including VGG-16, ResNet-56, and MobileNet following the setup in Section~\ref{sec:experiment_setup}. To simulate realistic deployment, each model contains six exits. We use the CIFAR-10 dataset to measure per-exit inference latency, and apply Kernel Density Estimation (KDE) to infer the exit index from the observed latency. Exit inference accuracy is computed as the fraction of samples whose inferred exit matches the ground-truth exit. %For each setting, we repeat the experiment for five times and report the average and standard deviation.

Table~\ref{tab:latencys} summarizes the results. The latency values across different exits are clearly distinguishable in both environments. The minimum latency gap between consecutive exits is 1.04 ms under TEE deployment and 0.26 ms under GPU deployment, both of which are sufficiently large relative to the observed standard deviations, confirming that latency is a reliable signal for exit index inference. The exit inference accuracy reported in Table~\ref{tab:latencys} is consistent with this observation, exceeding 99.8\% across all backbones under TEE deployment and 99.2\% under GPU deployment. These results confirm that the CSP can reliably recover the exit index from inference latency in both deployment settings, validating the feasibility of the end-to-end attack.

\subsection{Ablation Study} \label{sec:ablation}

\begin{table}[t]
\caption{Ablation study of the training objective on CIFAR-10 and VGGFace2S.  %``Full'' denotes the complete objective; other columns ablate individual components.
}
\centering
\setlength{\tabcolsep}{2pt}
\renewcommand{\arraystretch}{1.2}
\fontsize{8}{7}\selectfont
\begin{tabular}{ccccc}
\toprule

\textbf{Dataset}
&\textbf{Method}
& \textbf{Precision}
& \textbf{Recall}
& \textbf{F1 }
\\
\midrule

% cifar10-G
\multirow{5}{*}{\textbf{CIFAR-10}} 
&\textbf{Full}& 77.48\scalebox{0.8}{$\pm$9.25} & 84.98\scalebox{0.8}{$\pm$2.88} & 80.71\scalebox{0.8}{$\pm$5.44} \\

&\textbf{w/o $\boldsymbol{L^P}$}& 98.67\scalebox{0.8}{$\pm$2.98} \textcolor{blue}{$\uparrow$} & 0.80\scalebox{0.8}{$\pm$0.59} \textcolor{red}{$\downarrow$}  & 1.58\scalebox{0.8}{$\pm$1.16} \textcolor{red}{$\downarrow$} \\

&\textbf{w/o $\boldsymbol{L^T}$}& 47.96\scalebox{0.8}{$\pm$6.15} \textcolor{red}{$\downarrow$}& 81.96
\scalebox{0.8}{$\pm$2.68} \textcolor{red}{$\downarrow$} & 60.39\scalebox{0.8}{$\pm$5.46} \textcolor{red}{$\downarrow$}\\

&\textbf{w/o $\boldsymbol{L_{\text{smooth}}}$}  & 91.17\scalebox{0.8}{$\pm$1.81} \textcolor{blue}{$\uparrow$} & 41.64\scalebox{0.8}{$\pm$19.95} \textcolor{red}{$\downarrow$} & 54.76\scalebox{0.8}{$\pm$19.47} \textcolor{red}{$\downarrow$}\\

&\textbf{w/o $\boldsymbol{L_{\text{cov}}}$} & 66.67\scalebox{0.8}{$\pm$22.91} \textcolor{red}{$\downarrow$}& 54.10\scalebox{0.8}{$\pm$16.59} \textcolor{red}{$\downarrow$} & 57.82\scalebox{0.8}{$\pm$15.09} \textcolor{red}{$\downarrow$}\\

\midrule
% cifar10-SN
\multirow{5}{*}{\textbf{VGGFace2S}} 
&\textbf{Full} & 96.37\scalebox{0.8}{$\pm$0.67}& 79.48\scalebox{0.8}{$\pm$2.66}  & 87.08\scalebox{0.8}{$\pm$1.45} \\

&\textbf{w/o $\boldsymbol{L^P}$}& 80.00\scalebox{0.8}{$\pm$44.72} \textcolor{red}{$\downarrow$} & 0.39\scalebox{0.8}{$\pm$0.45} \textcolor{red}{$\downarrow$} & 0.77\scalebox{0.8}{$\pm$0.89} \textcolor{red}{$\downarrow$}\\

&\textbf{w/o $\boldsymbol{L^T}$}& 46.20\scalebox{0.8}{$\pm$5.83} \textcolor{red}{$\downarrow$}& 90.43\scalebox{0.8}{$\pm$0.77} \textcolor{blue}{$\uparrow$} & 60.96\scalebox{0.8}{$\pm$5.00} \textcolor{red}{$\downarrow$}\\

&\textbf{w/o $\boldsymbol{L_{\text{smooth}}}$}& 92.33\scalebox{0.8}{$\pm$4.93} \textcolor{red}{$\downarrow$} & 3.78\scalebox{0.8}{$\pm$2.82} \textcolor{red}{$\downarrow$} & 7.14\scalebox{0.8}{$\pm$5.16} \textcolor{red}{$\downarrow$}\\

&\textbf{w/o $\boldsymbol{L_{\text{cov}}}$}& 94.42\scalebox{0.8}{$\pm$1.05} \textcolor{red}{$\downarrow$}& 73.38\scalebox{0.8}{$\pm$4.04} \textcolor{red}{$\downarrow$}  & 83.27\scalebox{0.8}{$\pm$2.38} \textcolor{red}{$\downarrow$}\\

\bottomrule
\end{tabular}

\label{tab:ablation}
\end{table}

The training objective in Equation~\ref{eq:overall_loss} consists of three parts: the loss at exits before the designated exit $T$ ($L^P$), the loss at $T$ ($L^T$), and the standard cross-entropy loss at exits after $T$ ($L^A$). We ablate $L^P$ and $L^T$ to assess their individual contributions. Since removing either component entirely would cause severe accuracy degradation, we replace each with standard cross-entropy loss, yielding two ablation settings: \textbf{w/o $\boldsymbol{L^P}$} and \textbf{w/o $\boldsymbol{L^T}$}. Since both $L^P$ and $L^T$ are composed of selective label smoothing and entropy-covariance optimization, we ablate each of them separately, yielding two additional settings: \textbf{w/o $\boldsymbol{L}_{\text{smooth}}$} and \textbf{w/o $\boldsymbol{L}_{\text{cov}}$}. 
We compare all four ablation settings against the full loss on CIFAR-10 and VGGFace2S, using gray and white male as the target property, respectively.

\textbf{Experimental Results.} Results in table~\ref{tab:ablation} show that the full loss consistently achieves the best attack effectiveness. %, confirming that all components contribute to overall attack effectiveness. 
The full attack achieves the highest F1 score on both datasets. %, confirming that the full objective best balances precision and recall.
The full attack demonstrates superior precision, except in \textit{w/o $L_{\text{smooth}}$} and \textit{w/o $L^P$} on CIFAR-10, where recall drops substantially on ablated settings (from $84.98\%$ to $<42\%$).
The full attack achieves the best recall, except in \textit{w/o $L^T$} on VGGFace2S, where the model prioritizes recall over precision ($46.2\%$ precision versus $96.37\%$ for the full attack).

%The results align with our design rationale.\textit{w/o $L^T$} yields the lowest precision, while \textit{w/o $L^P$} yields the lowest recall. This aligns well with our design rationale that $L^T$ supports attack precision and $L^P$ improves attack recall. Overall, these results confirm that all components in our loss design are simultaneously necessary to achieve optimal attack performance. 

The results align with our design rationale. Removing $L^P$ most severely degrades recall (dropping to $\leq 0.8\%$ compared to $>79\%$ under the full objective), consistent with its role in preventing targeted inputs from exiting prematurely. Removing $L^T$ most severely degrades precision, consistent with its role in routing targeted samples to the designated exit. The component-level ablations further show that both selective label smoothing and entropy-covariance optimization contribute meaningfully, as removing either one degrades F1 score. Overall, these results confirm that all components of the training objective are necessary to achieve strong attack performance.

\subsection{Training Procedure Study} 
\label{sec:training_procedure_study}

\begin{table*}[t]
\caption{Attack effectiveness and stealthiness for different training schemes on CIFAR-10 and VGGFace2S.}
\centering
\renewcommand{\arraystretch}{1}  % 增加行高
\fontsize{8}{7}\selectfont

\begin{tabular}{ccccccc}
\toprule

\textbf{Dataset}
&\textbf{Method}
& \textbf{Precision (\%)}
& \textbf{Recall (\%)}
& \textbf{F1 ($10^{-2}$)}
& \textbf{Accuracy (\%)}
& \textbf{Speedup ($\times$)}
\\
\midrule
\multirow{2}{*}{\textbf{CIFAR-10}}
&\multirow{1}{*}{\textbf{Two-stage}}& \textbf{77.48\scalebox{0.8}{$\pm$9.25}} & 84.98\scalebox{0.8}{$\pm$2.88} &\textbf{80.71}\scalebox{0.8}{$\pm$5.44}&  \textbf{91.06\scalebox{0.8}{$\pm$0.30}}&\textbf{2.08 \scalebox{0.8}{$\pm$0.08}}\\
&\multirow{1}{*}{\textbf{Direct}}&67.77\scalebox{0.8}{$\pm$11.67} & \textbf{85.14\scalebox{0.8}{$\pm$2.78}} &75.02\scalebox{0.8}{$\pm$8.05}&  88.91\scalebox{0.8}{$\pm$0.42}&1.99 \scalebox{0.8}{$\pm$0.14}\\

\midrule

\multirow{2}{*}{\textbf{VGGFace2S}}
&\multirow{1}{*}{\textbf{Two-stage}} &\textbf{96.37}\scalebox{0.8}{$\pm$0.67}&\textbf{79.48}\scalebox{0.8}{$\pm$2.66}&\textbf{87.08}\scalebox{0.8}{$\pm$1.45}&\textbf{93.94}\scalebox{0.8}{$\pm$0.37}&\textbf{1.57}\scalebox{0.8}{$\pm$0.03}\\

&\multirow{1}{*}{\textbf{Direct}}&96.01\scalebox{0.8}{$\pm$0.93}&73.19\scalebox{0.8}{$\pm$2.37}&83.05\scalebox{0.8}{$\pm$1.74}&89.10\scalebox{0.8}{$\pm$2.34}&1.49\scalebox{0.8}{$\pm$0.09}\\

\bottomrule       
\end{tabular}

\label{tab:procedure}
\end{table*}

In Section~\ref{sec:loss-design-r3}, we propose a two-stage training pipeline that uses a clean multi-exit model as the starting point for attack training. In this section, we empirically demonstrate that this two-stage design achieves higher model accuracy compared to directly training with the attack objective from scratch.

We conduct experiments on CIFAR-10 and VGGFace2S, selecting gray and white male as the target properties, respectively. Table~\ref{tab:procedure} presents the results. The proposed two-stage approach (denoted as ``Two-Stage'' in Table~\ref{tab:procedure}) achieves better task accuracy than direct training with the adversarial objective (denoted as ``Direct''). Specifically, the two-stage method improves accuracy by 2.15\% on CIFAR-10 and 4.84\% on VGGFace2S compared to the direct method.

\subsection{Hyperparameter Study} \label{sec:hyperparameter}

In this section, we conduct hyperparameter studies on four aspects: the weighting hyperparameter $\lambda$, the label smoothing parameter $\epsilon$,  the choice of designated exit $T$, and the confidence threshold for early exiting.  %To obtain more stable results that isolate the impact of each hyperparameter, we deliberately reduce the variance introduced by the random number and location of exits. Specifically, unlike other results in this paper that are averaged over five runs with randomly chosen exit configurations, the experiments in this section use a fixed multi-exit model architecture and are repeated only across different random seeds.
All experiments are performed on CIFAR-10 and VGGFace2S,  using gray and white male as the target property, respectively.
To isolate the effect of each hyperparameter and reduce the uncertainty introduced by random exit configurations, we fix the multi-exit model architecture across runs and vary only the random seed, unlike the main experiments where the number and location of exits are randomly sampled across five independent trials.

%All experiments are performed on CIFAR-10 and VGGFace2S,  using gray and white male as the target property, respectively. Details are in Appendix~\ref{sec:hyperparameter_details}.

% We conduct hyperparameter studies on four aspects: the label smoothing parameter $\epsilon$ (varied from 0.1 to 1), the weighting hyperparameter $\lambda$ (varied from 0.5 to 5.0), the choice of designated exit $T$ (ranging from 15\% to 90\% of total layers), and the confidence threshold for early exiting (varied from 0.5 to 0.99). All experiments are performed on CIFAR-10 and VGGFace2S,  using gray and white male as the target property, respectively. Details are in Appendix~\ref{sec:hyperparameter_details}.

% \textbf{Experimental Results.} Our attack demonstrates robustness across a wide range of hyperparameter settings. For $\epsilon$, the attack performs well across $[0.1, 0.6]$; for $\lambda$, a moderate range of $[1.5, 2.5]$ achieves a good balance between effectiveness and stealthiness; for the designated exit, deeper layers (beyond 75\% of total layers) yield stronger attack performance while preserving stealthiness; for the confidence threshold, the attack remains robust from 0.5 to 0.99, maintaining high precision and recall with minimal accuracy degradation. Detailed results and figures are provided in Appendix~\ref{sec:hyperparameter_details}.

\begin{figure*}[t]
\centering
\includegraphics[width=0.99\linewidth]{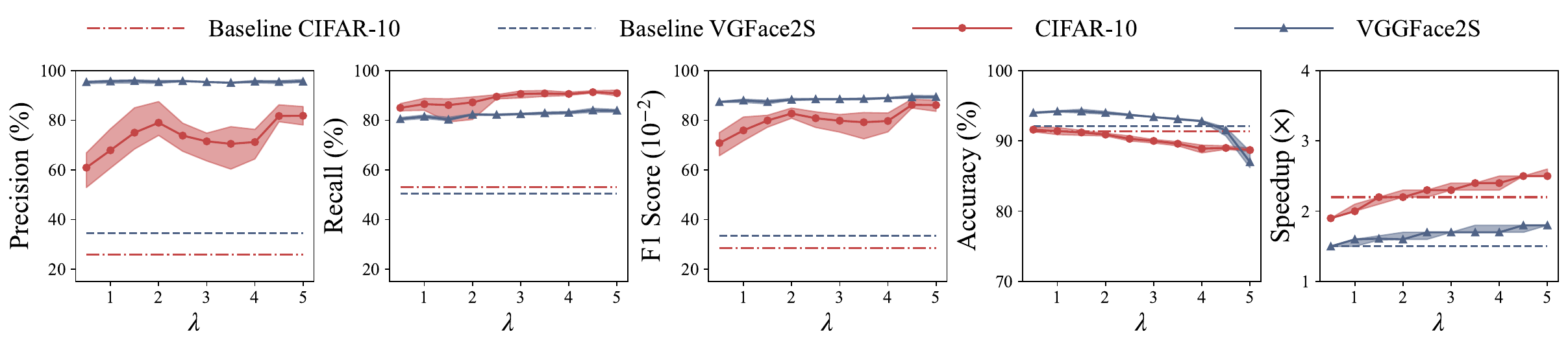}
\caption{Attack effectiveness and stealthiness %under different weighting hyperparameters $\lambda$. 
under different $\lambda$ values. We report the best results of the baselines for each setting in the baseline curves.
}
\label{fig:hyperparameter_cov}
\end{figure*}

\begin{figure*}[t]
\centering
\includegraphics[width=0.99\linewidth]{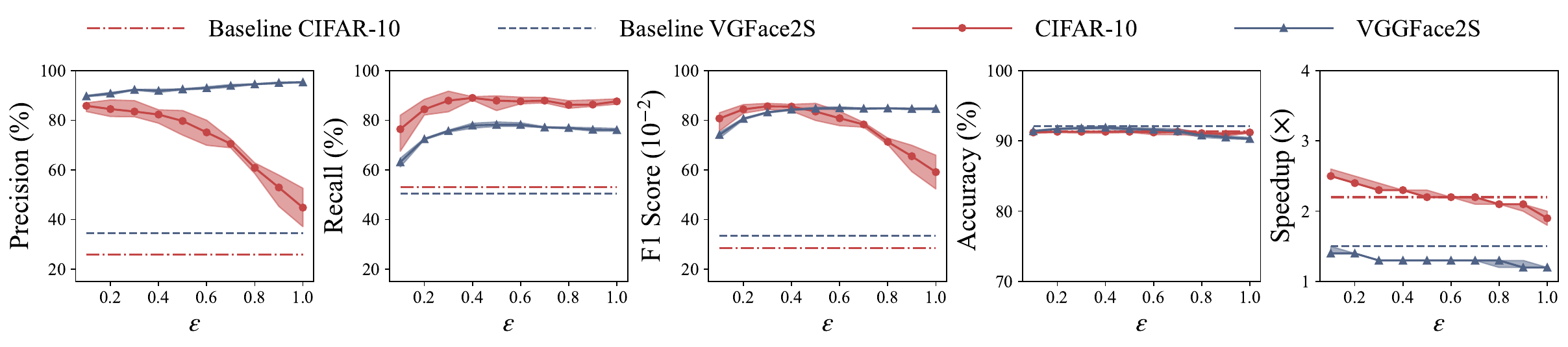}
\caption{Attack effectiveness and stealthiness %under different label smoothing parameter $\epsilon$.
under different $\epsilon$ values. We report the best results of the baselines for each setting in the baseline curves.
}
\label{fig:hyperparameter_smooth}
\end{figure*}

\subsubsection{Impact of Weighting Hyperparameter}

We study the impact of $\lambda$ in Equations~\ref{eq:target_loss} and~\ref{eq:before_loss}, which balances the two loss terms $L_{\text{smooth}}$ and $L_{\text{cov}}$. In our experiments, we vary $\lambda$ from 0.5 to 5.0 in increments of 0.5. Unless otherwise specified, all hyperparameter studies are conducted on CIFAR-10 and VGGFace2S with gray and white male as the target property, respectively.
%conducting experiments on CIFAR-10 and VGGFace2S with gray and white male as the target properties, respectively. All hyperparameter studies in this appendix are conducted under these tasks. 

The results are in Figure~\ref{fig:hyperparameter_cov}, showing a clear trade-off between main task accuracy and other metrics. As $\lambda$ increases, attack effectiveness (precision, recall, and F1 score) generally improves. This is expected, as a larger $\lambda$ imposes a stronger constraint on the confidence distributions at each exit, widening the gap between targeted and non-targeted inputs and making exit routing a more reliable indicator of the target property. However, an excessively large $\lambda$ interferes with the learning of the main task, leading to noticeable accuracy degradation. Inference speedup shows an increasing trend with $\lambda$, which we attribute to the larger confidence difference between targeted and non-targeted inputs before the designated exit, causing more non-targeted inputs to exit earlier on average.

Despite these trade-offs, an appropriate range of $\lambda$ can be identified. For $\lambda \in [1.5, 2.5]$, precision, recall, F1 score, and inference speedup remain consistently above 73\%, 80\%, 0.8, and 1.6$\times$, respectively, with accuracy degradation below 1\%. The attack is therefore not sensitive to the precise choice of $\lambda$ within this range, and attackers can set this parameter without extensive tuning.

\subsubsection{Impact of Label Smoothing Parameter} 

We study the impact of the label smoothing parameter $\epsilon$ introduced in Section~\ref{sec:loss-design-r1} by varying it from 0.1 to 1.0 in steps of 0.1.

Figure~\ref{fig:hyperparameter_smooth} shows the results. Our attack performs well across a broad range of $\epsilon$. For $\epsilon \in [0.1, 0.6]$, precision, recall, and F1 score are consistently above 75\%, 63\%, and 0.74, respectively, substantially outperforming the best baseline values (below 36\%, 54\%, and 0.35). Recall generally increases with $\epsilon$, as a larger smoothing parameter can more effectively suppress the confidence of targeted inputs at exits before the designated exit $T$, reducing premature termination and improving recall. F1 score remains relatively stable within this range, with differences below 0.1. Precision decreases on CIFAR-10 but increases on VGGFace2S as $\epsilon$ increases. We leave a detailed investigation on precision to future work. Model accuracy remains stable for $\epsilon \in [0.1, 0.6]$, with differences below 1\%. Inference speedup decreases gradually as $\epsilon$ increases, but the degradation remains within 0.5 across this range.

\begin{figure*}[t]
\centering
\includegraphics[width=0.99\linewidth]{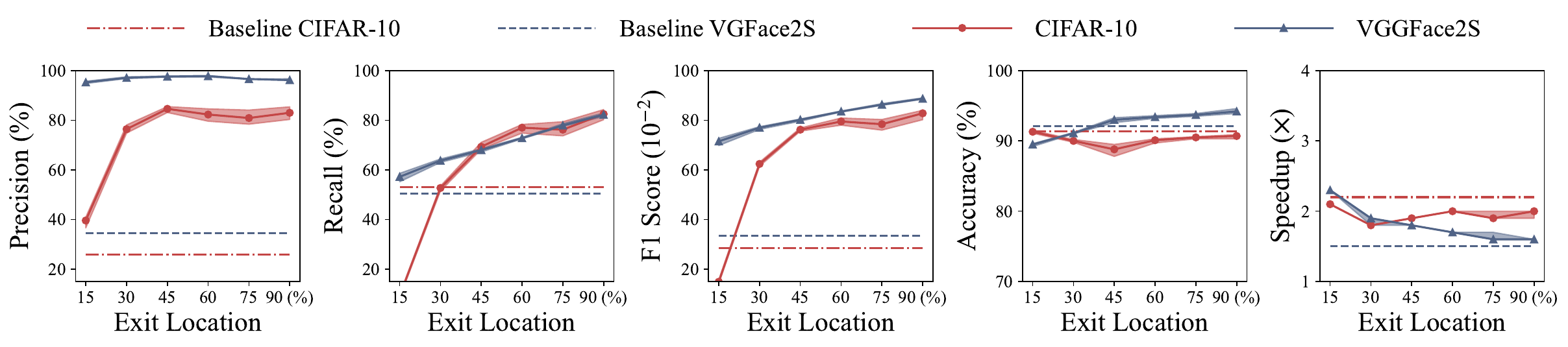}
\caption{Attack effectiveness and stealthiness under different designated exits ($T$).
We report the best results of the baselines for each setting in the baseline curves.}
\label{fig:hyperparameter_location}
\end{figure*}

\begin{figure*}[t]
\centering
\includegraphics[width=0.99\linewidth]{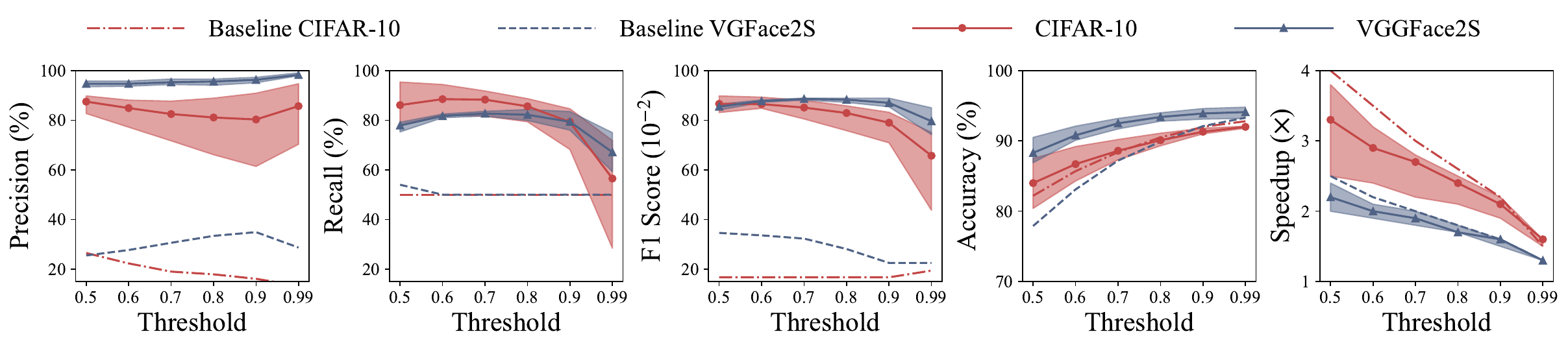}
\caption{Attack effectiveness and stealthiness under different confidence thresholds. We report the best results of the baselines for each setting in the baseline curves. Since the confidence threshold affects the attack effectiveness of the FP and LR baselines and the attack stealthiness of all baselines, the baseline curves are dynamic rather than linear.}
\label{fig:thres}
\end{figure*}

\subsubsection{Impact of Choice of Designated Exit} \label{sec:exit_impact}

In the main experiments, we simply choose the exit immediately preceding the final exit as the designated exit. In this section, we investigate the impact of this choice. We select six exit positions corresponding to 15\%, 30\%, 45\%, 60\%, 75\%, and 90\% of the total layers and use each as the designated exit, resulting in six different experimental settings.

The results are shown in Figure~\ref{fig:hyperparameter_location}. We observe that deeper exits generally yield better attack effectiveness while maintaining stealthiness. For instance, when the designated exit corresponds to $\geq 75\%$ of the total layers, our attack achieves precision, recall and F1 scores above 80\%, 76\%, and 0.78, respectively, on both datasets, while maintaining model accuracy above 90\% for CIFAR-10 and 93\% for VGGFace2S, with inference speedup above $1.6\times$.

Regarding attack effectiveness, precision, recall, and F1 score generally increase as the designated exit goes deeper. A possible explanation is that deeper exits have access to richer representations, allowing the model to better distinguish between targeted and non-targeted inputs. 

Regarding attack stealthiness, the behavior differs across datasets. On CIFAR-10, model accuracy and inference speedup first decrease and then increase as the exit goes deeper. This is because many samples on CIFAR-10 are naturally routed to intermediate exits in a clean model, and designating one of these exits for attack purposes disrupts their routing, reducing both accuracy and speedup. On VGGFace2S, model accuracy increases while inference speedup decreases as the exit goes deeper. A possible explanation is that the face classification task is more challenging than image classification on CIFAR-10, and manipulating deeper layers eases the learning task at the cost of slightly lower speedup.

\subsubsection{Impact of Confidence Threshold}

In the main experiments, we use a confidence threshold of 0.9 for early exiting. In this section, we conduct experiments with varying thresholds. We select thresholds of 0.5, 0.6, 0.7, 0.8, 0.9, and 0.99, resulting in six settings in total. We note that thresholds below 0.5 lead to excessively low model accuracy and are therefore not considered. We also exclude a threshold of 1.0, since it would effectively prevent samples from exiting at internal exits. Instead, we use 0.99 as the upper bound.

Figure~\ref{fig:thres} shows the results. Our attack remains robust across all evaluated thresholds. Specifically, precision exceeds 80\% for all settings, significantly higher than that of the baselines (below 36\%). Recall is above 67\% under all settings except the 0.99 threshold on CIFAR-10, which still outperforms the best baseline (57\%). F1 scores exceed 0.65 for our attack compared to below 0.35 for the baselines. Meanwhile, the drop in task accuracy remains within 0.8\%, and inference speedup decreases by at most 0.7 across all settings.

Although the results demonstrate strong attack effectiveness across different thresholds, a clear trade-off exists between model accuracy and inference speedup. As the threshold increases, the model achieves better task accuracy but lower inference speedup. This aligns with the inherent behavior of multi-exit models: lower thresholds enable more samples to exit at earlier layers, resulting in higher speedup, but may also cause earlier exits to output lower-confidence predictions, thereby reducing model accuracy. In practice, a relatively high threshold is more appropriate to preserve model utility, and our default choice of 0.9 reflects this consideration.

\subsection{Inferring Multiple Properties Simultaneously} \label{sec:multi_property} %Simultaneously}

The main experiments focus on a single target property. Our method can extend naturally to the simultaneous inference of multiple mutually exclusive properties. %, where each input possesses at most one target property.
Specifically, we assign each target property $p$ to a distinct designated exit $T_p$. For the loss $L^{T}$ at exit $T_p$, we treat inputs possessing property $p$ as targeted and all other inputs (including those with other target properties) as non-targeted, 
ensuring that each designated exit learns to produce high confidence specifically for its assigned property.
%This ensures that each designated exit learns to produce high confidence specifically for its assigned property. 
For the loss $L^{P}$ at exits before the designated exits, we treat inputs with any target property as targeted and all remaining inputs as non-targeted, preventing any targeted input from exiting prematurely.
%all inputs that possess any target property as targeted, and  samples that possess none of the target properties as non-targeted. This prevents samples with any target property from exiting early, pushing them deeper to reach their respective designated exits.

Using VGGFace2S with VGG-16, we select three mutually exclusive properties: White Male, Non-White Female, and Non-White Male. We assign  the last three internal exits as their designated exits, respectively. 
%Table~\ref{tab:multi} presents the experimental results, our attack significantly outperforms the baselines. 
As shown in Table~\ref{tab:multi}, our attack substantially outperforms the baselines: precision exceeds 83\% (versus below 37\%), recall exceeds 60\% (versus below 57\%), and F1 score exceeds 0.71 (versus below 0.285). Model accuracy is unaffected and inference speedup decreases by 0.04, confirming that the extension preserves stealthiness.

\begin{table}[t]
\caption{Attack performance for simultaneously inferring three properties on VGGFace2S using VGG-16
}
\centering
\setlength{\tabcolsep}{1.5pt}
\renewcommand{\arraystretch}{1}  % 增加行高
% \small % 使用small字体
\fontsize{7}{7}\selectfont
\begin{tabular}{ccccccc}
\toprule

\textbf{Property}
&\textbf{Method}
& \textbf{Precision}
& \textbf{Recall}
& \textbf{F1 ($10^{-2}$)}
& \textbf{Accuracy}
& \textbf{Speedup}
\\
\midrule
\multirow{4}{*}{\textbf{White Male}}
&\multirow{1}{*}{\textbf{RG}}& 14.50\scalebox{0.8}{$\pm$0.00}&50.00\scalebox{0.8}{$\pm$0.00}&22.48\scalebox{0.8}{$\pm$0.00}&91.80\scalebox{0.8}{$\pm$0.27}&\textbf{1.51\scalebox{0.8}{$\pm$0.05}}\\
&\multirow{1}{*}{\textbf{FP}}& 36.45\scalebox{0.8}{$\pm$4.83}&15.86\scalebox{0.8}{$\pm$6.04}&21.26\scalebox{0.8}{$\pm$6.58}&91.80\scalebox{0.8}{$\pm$0.27}&\textbf{1.51\scalebox{0.8}{$\pm$0.05}}\\
&\multirow{1}{*}{\textbf{LR}} &20.32\scalebox{0.8}{$\pm$0.88}&50.40\scalebox{0.8}{$\pm$12.63}&28.44\scalebox{0.8}{$\pm$3.02}&91.80\scalebox{0.8}{$\pm$0.27}&\textbf{1.51\scalebox{0.8}{$\pm$0.05}}\\
&\multirow{1}{*}{\textbf{Our}}& \textbf{97.73\scalebox{0.8}{$\pm$0.34}}&\textbf{72.48\scalebox{0.8}{$\pm$1.64}}&\textbf{83.22\scalebox{0.8}{$\pm$0.96}}&\textbf{93.79\scalebox{0.8}{$\pm$0.42}}&1.47\scalebox{0.8}{$\pm$0.07}\\
\midrule

\multirow{4}{*}{\textbf{\shortstack{\textbf{Non-White} \\ \textbf{Female}}}}
&\multirow{1}{*}{\textbf{RG}}& 14.06\scalebox{0.8}{$\pm$0.00}&50.00\scalebox{0.8}{$\pm$0.00}&21.95\scalebox{0.8}{$\pm$0.00}&91.80\scalebox{0.8}{$\pm$0.27}&\textbf{1.51\scalebox{0.8}{$\pm$0.05}}\\
&\multirow{1}{*}{\textbf{FP}}&21.45\scalebox{0.8}{$\pm$1.62}&14.77\scalebox{0.8}{$\pm$6.43}&16.80\scalebox{0.8}{$\pm$4.18}&91.80\scalebox{0.8}{$\pm$0.27}&\textbf{1.51\scalebox{0.8}{$\pm$0.05}}\\
&\multirow{1}{*}{\textbf{LR}}&15.83\scalebox{0.8}{$\pm$0.33}&50.64\scalebox{0.8}{$\pm$12.48}&23.88\scalebox{0.8}{$\pm$1.79}&91.80\scalebox{0.8}{$\pm$0.27}&\textbf{1.51\scalebox{0.8}{$\pm$0.05}}\\
&\multirow{1}{*}{\textbf{Our}}& \textbf{92.71\scalebox{0.8}{$\pm$0.32}}&\textbf{60.66\scalebox{0.8}{$\pm$3.93}}&\textbf{73.26\scalebox{0.8}{$\pm$2.78}}&\textbf{93.79\scalebox{0.8}{$\pm$0.42}}&1.47\scalebox{0.8}{$\pm$0.07}\\
\midrule
\multirow{4}{*}{\textbf{\shortstack{\textbf{Non-White} \\ \textbf{Male}}}}
&\multirow{1}{*}{\textbf{RG}}& 7.54\scalebox{0.8}{$\pm$0.00}&50.00\scalebox{0.8}{$\pm$0.00}&13.10\scalebox{0.8}{$\pm$0.00}&91.80\scalebox{0.8}{$\pm$0.27}&\textbf{1.51\scalebox{0.8}{$\pm$0.05}}\\
&\multirow{1}{*}{\textbf{FP}}&13.30\scalebox{0.8}{$\pm$1.05}&19.10\scalebox{0.8}{$\pm$5.89}&15.33\scalebox{0.8}{$\pm$2.05}&91.80\scalebox{0.8}{$\pm$0.27}&\textbf{1.51\scalebox{0.8}{$\pm$0.05}}\\
&\multirow{1}{*}{\textbf{LR}}&9.83\scalebox{0.8}{$\pm$0.75}&56.87\scalebox{0.8}{$\pm$10.92}&16.62\scalebox{0.8}{$\pm$1.00}&91.80\scalebox{0.8}{$\pm$0.27}&\textbf{1.51\scalebox{0.8}{$\pm$0.05}}\\
&\multirow{1}{*}{\textbf{Our}}& \textbf{83.64\scalebox{0.8}{$\pm$1.40}}&\textbf{62.33\scalebox{0.8}{$\pm$1.13}}&\textbf{71.42\scalebox{0.8}{$\pm$1.07}}&\textbf{93.79\scalebox{0.8}{$\pm$0.42}}&1.47\scalebox{0.8}{$\pm$0.07}\\
\bottomrule       
\end{tabular}

\label{tab:multi}
\end{table}

\subsection{Extended Attack Scenario}
\label{sec:attacker3}
%In the main experiments, users deploy the complete multi-exit model provided by the CSP. We further consider a relaxed scenario, also studied in~\cite{kaya2019shallow,dong2023mind,chen2023dark}, where the CSP provides only a pretrained backbone and users independently insert internal classifiers. We show that the CSP can still mount the attack by manipulating the backbone alone; the technical details of the adapted training procedure are provided in Appendix~\ref{sec:attacker3_detail}.

%We conduct experiments on CIFAR-10 and VGGFace2S with VGG-16, targeting gray and white male respectively, simulating user exit insertion by reusing the setup of Section~\ref{sec:experiment_setup}. As shown in Appendix~\ref{sec:attacker3_detail}, our attack substantially outperforms the baselines: precision exceeds 64\% (versus below 35\%), recall exceeds 54\% (versus below 50\%), and F1 exceeds 0.59 (versus below 0.22). Model accuracy drops by at most 2\% and inference speedup decreases by less than 0.3. %While effectiveness is slightly lower than in the main setting,  The results confirm that the attack generalizes to this more challenging scenario.

\begin{table}[h]
\caption{Attack effectiveness and stealthiness under the extended attack scenario with VGG16.}
\centering
\setlength{\tabcolsep}{1.5pt}
\renewcommand{\arraystretch}{1}  % 增加行高
% \small  % 使用small字体
\fontsize{7}{7}\selectfont

\begin{tabular}{ccccccc}
\toprule

\textbf{Dataset}
&\textbf{Method}
& \textbf{Precision}
& \textbf{Recall}
& \textbf{F1 ($10^{-2}$)}
& \textbf{Accuracy}
& \textbf{Speedup}
\\
\midrule
\multirow{4}{*}{\textbf{CIFAR-10}}
& \textbf{RG} & 10.00\scalebox{0.8}{$\pm$0.00}&50.00\scalebox{0.8}{$\pm$0.00}&16.67\scalebox{0.8}{$\pm$0.00}&\textbf{91.27}\scalebox{0.8}{$\pm$0.23}&2.16\scalebox{0.8}{$\pm$0.07} \\

& \textbf{FP} & 16.53\scalebox{0.8}{$\pm$0.85}&9.66\scalebox{0.8}{$\pm$3.19}&11.89\scalebox{0.8}{$\pm$2.44}&\textbf{91.27}\scalebox{0.8}{$\pm$0.23}&2.16\scalebox{0.8}{$\pm$0.07} \\

& \textbf{LR} & 25.75\scalebox{0.8}{$\pm$6.39}&53.06\scalebox{0.8}{$\pm$9.73}&33.33\scalebox{0.8}{$\pm$4.62}&\textbf{91.27}\scalebox{0.8}{$\pm$0.23}&2.16\scalebox{0.8}{$\pm$0.07} \\
&\multirow{1}{*}{\textbf{Our}}& \textbf{64.42}\scalebox{0.8}{$\pm$10.29}&\textbf{69.86}\scalebox{0.8}{$\pm$7.08}&\textbf{66.17}\scalebox{0.8}{$\pm$3.21}& 89.83\scalebox{0.8}{$\pm$0.20}&\textbf{2.17}\scalebox{0.8}{$\pm$0.09}\\
\midrule

\multirow{4}{*}{\textbf{VGGFace2S}}
& \textbf{RG} &14.50\scalebox{0.8}{$\pm$0.00}&50.00\scalebox{0.8}{$\pm$0.00}&22.48\scalebox{0.8}{$\pm$0.00}&\textbf{91.80}\scalebox{0.8}{$\pm$0.27}&\textbf{1.51}\scalebox{0.8}{$\pm$0.05}\\
& \textbf{FP} &36.45\scalebox{0.8}{$\pm$4.83}&15.86\scalebox{0.8}{$\pm$6.04}&21.26\scalebox{0.8}{$\pm$6.58}&\textbf{91.80}\scalebox{0.8}{$\pm$0.27}&\textbf{1.51}\scalebox{0.8}{$\pm$0.05}\\
& \textbf{LR} &20.32\scalebox{0.8}{$\pm$0.88}&50.40\scalebox{0.8}{$\pm$12.63}&28.44\scalebox{0.8}{$\pm$3.02}&\textbf{91.80}\scalebox{0.8}{$\pm$0.27}&\textbf{1.51}\scalebox{0.8}{$\pm$0.05}\\
&\multirow{1}{*}{\textbf{Our}}&\textbf{69.62}\scalebox{0.8}{$\pm$17.78}&\textbf{54.32}\scalebox{0.8}{$\pm$5.20}&\textbf{59.45}\scalebox{0.8}{$\pm$5.97}& 90.10\scalebox{0.8}{$\pm$0.14}& 1.19\scalebox{0.8}{$\pm$0.04}\\

\bottomrule       
\end{tabular}

\label{tab:attacker3}
\end{table}

In the main experiments, we assume that users directly deploy the multi-exit model provided by the CSP. In this section, we relax this assumption and consider an alternative scenario, also studied in~\cite{kaya2019shallow,dong2023mind,chen2023dark}: the CSP provides only a pre-trained backbone model, and users independently insert internal classifiers to create multi-exit models for efficient inference. We demonstrate that the CSP can still launch the input privacy inference attack by manipulating only the backbone.

In this scenario, the CSP cannot know in advance where users will insert exits, but can estimate a possible range of layers where at least one exit is likely to be inserted. For instance, based on a survey of existing designs~\cite{kaya2019shallow,dong2023mind,chen2023dark}, we observe that at least one internal exit is located in the latter 40\% of layers for most designs.

For attack training, we insert exits at all potential positions in the backbone and modify the loss defined in Equation~\ref{eq:overall_loss}. Our key idea is to treat all exits within the estimated range as designated exits. Specifically, for exits inside the estimated range, we apply Equation~\ref{eq:target_loss}; for exits before the range, we use Equation~\ref{eq:before_loss}; and for exits after the range, we use Equation~\ref{eq:after_loss}.
For the training pipeline, we first train clean multi-exit models, then freeze the internal classifiers and fine-tune only the backbone parameters using the modified loss (see Section~\ref{sec:loss-design-r3}). After attack training, the CSP releases only the backbone part of the trained multi-exit model.

We conduct experiments on CIFAR-10 and VGGFace2S using the VGG-16 backbone, targeting the properties gray and white male, respectively. We set the estimated range for user-inserted exit(s) to the final 40\% of layers during attack training. We simulate the user's exit insertion process by reusing the setup from Section~\ref{sec:experiment_setup}. We first randomly select a total number of exits from $[4, 6]$, then insert them at randomly chosen positions. We repeat the insertion process five times and report the average and standard deviation of the results. We note that it is possible that no exits are inserted in the latter 40\% of layers, in which case the precision, recall, and F1 score are all zero.

Table~\ref{tab:attacker3} summarizes the experimental results. Our attack shows strong effectiveness, significantly outperforming the baselines. It achieves attack precision above 64\%, compared to below 36.5\% for the baselines. Our attack also improves Recall from below 53.1\% to exceeding 54\%. Consequently, the F1 score increases from below 0.34 to above 0.59. Model accuracy and inference speedup are also preserved. The accuracy drop is within 2\%, and the degradation in inference speedup is less than 0.32.

Although attack effectiveness slightly decreases compared to Section~\ref{sec:main_results}, this relaxed setting still achieves strong results and demonstrates that our attack design generalizes well across different attack scenarios.

%另一种使用多出口网络的方式\cite{kaya2019shallow,dong2023mind,chen2023dark}，是基于pretrained的backbone，插入并训练内部分类器。在这种场景下，我们假设CSP提供者仅仅提供backbone，训练则由用户自行完成。为了成功实现这种场景下的攻击，我们需要对原方案进行两点修改。1.CSP提供者无法得知用户将会在什么位置插入出口，但是可以通过猜测用户可能插入的区间$[T_{start},T_{end}]$。因此，在训练时我们需要将所有可能的位置均插入出口，对于小于$T_{start}$的出口，使用\ref{eq:before_loss}进行训练，对于$[T_{start},T_{end}]$内的出口，使用\ref{eq:target_loss}进行训练,对于大于$T_{end}]$的出口，使用\ref{eq:after_loss}进行训练。2.CSP提供者只能够发布backbone，因此在微调的过程中，冻结干净的内部分类器，仅仅训练主干的参数。基于改进后的方案，我们在CIFAR-10，目标属性grey和VGGface，目标属性white male上使用VGG-16进行了实验，$[T_{start},T_{end}]$为后百分之40\%的层。实验结果展示在\ref{tab:attacker3}中,相较于CSP提供完全的多出口网络情况，攻击效果变差，但仍然显著优于baseline（precision分别为随机猜测的6.4倍和4.8倍），Accuracy的下降控制在2\%以内，Operations的上升控制在30\%以内。

\section{Possible Defenses}

% To mitigate the privacy leakage introduced by our attack, we propose potential defense strategies. Recall that the attack relies on inference latency to infer the exit index and thereby determine the target property. Therefore, a natural defense is to introduce interference in the inference latency.

% One approach is to randomize the early exiting behavior. For instance, instead of exiting immediately when the confidence threshold is reached, a sample could exit with probability $p$~\cite{warner1965randomized}. Another approach is to add random dummy computations at each exit, varying the inference time and making it difficult for an attacker to accurately infer the exit index from latency measurements.

% However, these schemes inevitably introduce additional time overhead, undermining the speedup benefits of multi-exit models. Thus, how to protect privacy without compromising inference efficiency remains an open problem for future research.

A natural defense against our attack is to obscure inference latency, thereby preventing the CSP from inferring the exit index. Based on this principle, we propose two possible defensive methods. The first method is to randomize the early exit decision \textbf{(Randomized Exit)}. At each intermediate exit, instead of exiting deterministically once the confidence threshold is reached, a sample exits with a probability $p$~\cite{warner1965randomized}. This mechanism introduces uncertainty into the actual exit position. The second method is to insert random dummy computations at each exit to obscure the relationship between latency and exit depth \textbf{(Dummy Computation)}. After a sample exits, additional randomized computation is introduced to weaken the strong correlation between its inference latency and exit index. 

We conduct experiments to evaluate the effectiveness of these possible defenses. All experiments are conducted on the CIFAR-10 dataset, with gray as the target property. For the randomized exit defense, we vary $p$ from $0.1$ to $1$ and report the attack F1 score and the inference speedup. For the dummy computation defense, since the exact form of dummy computation may vary across deployment environments, we evaluate this defense through simulation. Specifically, we first draw a value $\Delta t$ from a Gaussian distribution $\mathcal{N}(0,\sigma^2)$. We then set the latency observed by the CSP as the original inference latency plus the absolute value of $\Delta t$. We vary $\sigma$ from $0$ to $1\times10^{-3}$ and report the results.

\begin{figure}[t]
\centering
\includegraphics[width=1\linewidth]{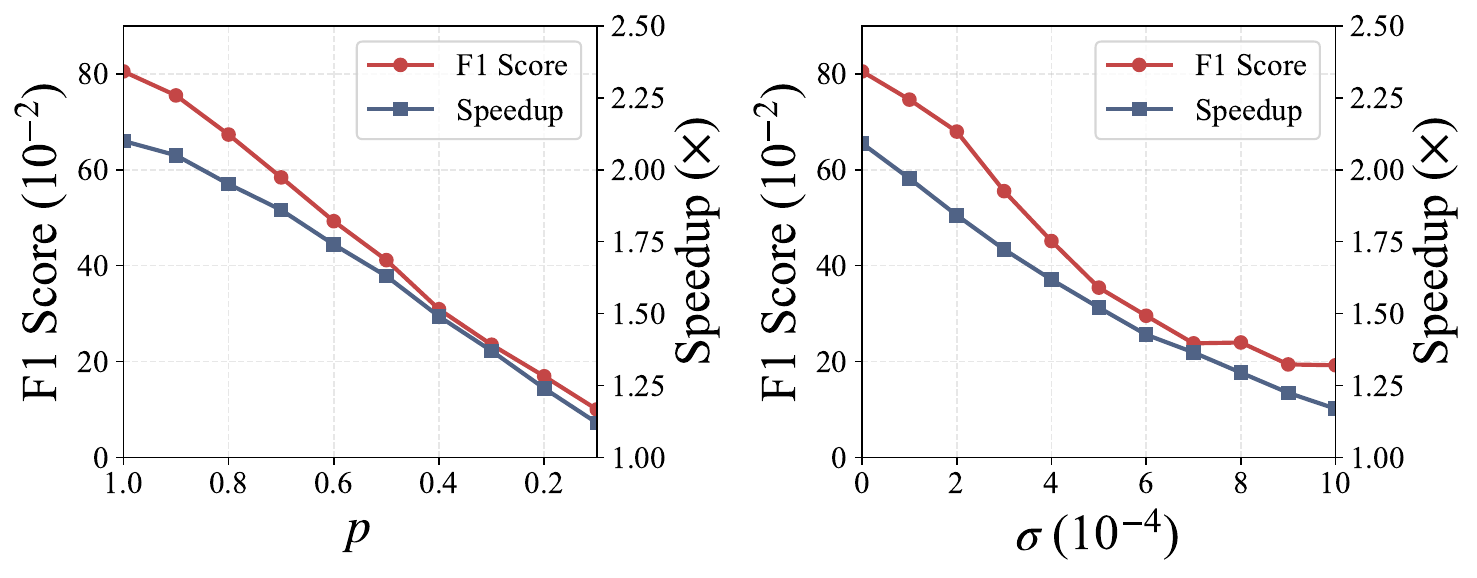}
\caption{Attack effectiveness and inference speedup under randomized exit defense and dummy computation defense}
\label{fig:defense_pic}
\end{figure}

As shown in figure~\ref{fig:defense_pic}, the attack F1 score and inference speedup decrease simultaneously as $p$ decreases or $\sigma$ increases. 
These results demonstrate the effectiveness of both defense mechanisms in mitigating latency-based privacy leakage. However, this protection inevitably weakens the acceleration benefits of multi-exit networks. Therefore, achieving effective privacy protection without significantly compromising inference efficiency remains an important direction for future research.

Other defenses could also be explored. Fine-tuning the model on trusted data may weaken the malicious routing behavior. Inspecting the model parameters before deployment may help detect the attack. We leave the evaluation of these methods to future work.

% \ynote{todo rephrase}
% A promising direction is anomaly detection based on model parameters: specifically, a specially fine-tuned model may exhibit parameter-level differences from a normal model, and detecting such differences would not affect inference efficiency. We will investigate this further in future work.

\section{Conclusion}
%In this paper, we design a privacy attack targeting multi-exit models. On a manipulated multi-exit model, a malicious cloud service provider can infer sensitive property of user inputs solely by observing inference time. Experimental results demonstrate that this attack can effectively achieve without compromising model accuracy or inference efficiency. This poses new challenges for the security of cloud-based AI services.

We identified a new privacy vulnerability in multi-exit networks: a malicious cloud service provider can infer sensitive properties of user inputs solely from inference latency, by supplying a manipulated model that routes targeted inputs to a designated exit, without accessing model inputs, internal states, or model outputs. 
We proposed a training framework combining selective label smoothing and entropy-covariance optimization to realize this attack. %Experiments across diverse datasets, architectures, and target properties demonstrate strong effectiveness and stealthiness. 
Extensive experiments across diverse settings demonstrate strong attack effectiveness and stealthiness. 
Our findings highlight the privacy risks of dynamic inference mechanisms, underscoring the need for defenses that protect privacy without sacrificing efficiency.

\bibliographystyle{IEEEtran}
\bibliography{property_inference}

\end{document}